\documentclass[letterpaper]{article} 
\usepackage{aaai2026}
\usepackage{times}  
\usepackage{helvet}  
\usepackage{courier}  
\usepackage[hyphens]{url}  
\usepackage{graphicx} 
\usepackage{natbib}  
\usepackage{caption} 
\usepackage{algorithm}
\usepackage{algorithmic}

\usepackage{booktabs}
\usepackage{tabularx}
\usepackage{makecell}
\usepackage{colortbl}
\usepackage[most]{tcolorbox}
\usepackage{bm}
\usepackage{multirow}
\usepackage{threeparttable}
\usepackage{rotating}

\definecolor{dark-red}{RGB}{255,0,0}
\definecolor{dark-green}{RGB}{0,200,0}
\definecolor{theme-color}{RGB}{153,153,255}

\usepackage{newfloat}
\usepackage{listings}
\DeclareCaptionStyle{ruled}{labelfont=normalfont,labelsep=colon,strut=off} 
\floatstyle{ruled}
\newfloat{listing}{tb}{lst}{}
\floatname{listing}{Listing}
\newcommand{\name}{\textsc{R2Vul}}

\nocopyright

\title{$\name$: Learning to Reason about Software Vulnerabilities with Reinforcement Learning and Structured Reasoning Distillation}
\author{
    Martin Weyssow\textsuperscript{\rm 1}\thanks{Corresponding author.}, Chengran Yang\textsuperscript{\rm 1}, Junkai Chen\textsuperscript{\rm 1}, Ratnadira Widyasari\textsuperscript{\rm 1}, Ting Zhang\textsuperscript{\rm 1}, Huihui Huang\textsuperscript{\rm 1}, Huu Hung Nguyen\textsuperscript{\rm 1}, Yan Naing Tun\textsuperscript{\rm 1}, Tan Bui\textsuperscript{\rm 1}, Yikun Li\textsuperscript{\rm 1}, Ang Han Wei\textsuperscript{\rm 2}, Frank Liauw\textsuperscript{\rm 2}, Eng Lieh Ouh\textsuperscript{\rm 1}, Lwin Khin Shar\textsuperscript{\rm 1}, David Lo\textsuperscript{\rm 1}
}
\affiliations{
    \textsuperscript{\rm 1}Singapore Management University\\
    \textsuperscript{\rm 2}GovTech Singapore\\

    mweyssow@smu.edu.sg
}

\usepackage{bibentry}

\newcounter{checksubsection}
\newcounter{checkitem}[checksubsection]

\newcommand{\checksubsection}[1]{%
  \refstepcounter{checksubsection}%
  \paragraph{\arabic{checksubsection}. #1}%
  \setcounter{checkitem}{0}%
}

\newcommand{\checkitem}{%
  \refstepcounter{checkitem}%
  \item[\arabic{checksubsection}.\arabic{checkitem}.]%
}
\newcommand{\question}[2]{\normalcolor\checkitem #1 #2 \color{blue}}
\newcommand{\ifyespoints}[1]{\makebox[0pt][l]{\hspace{-15pt}\normalcolor #1}}

\begin{document}

\maketitle

\begin{abstract}

Large language models (LLMs) have shown promising performance in software vulnerability detection, yet their reasoning capabilities remain unreliable.
We propose $\name$, a method that combines reinforcement learning from AI feedback (RLAIF) and structured reasoning distillation to teach small code LLMs to detect vulnerabilities while generating security-aware explanations. 
Unlike prior chain-of-thought and instruction tuning approaches, $\name$ rewards well-founded over deceptively plausible vulnerability explanations through RLAIF, which results in more precise detection and high-quality reasoning generation.
To support RLAIF, we construct the first multilingual preference dataset for vulnerability detection, comprising 18,000 high-quality samples in C\#, JavaScript, Java, Python, and C.
We evaluate $\name$ across five programming languages and against four static analysis tools, eight state-of-the-art LLM-based baselines, and various fine-tuning approaches.
Our results demonstrate that a 1.5B $\name$ model exceeds the performance of its 32B teacher model and leading commercial LLMs such as Claude-4-Opus.
Furthermore, we introduce a lightweight calibration step that reduces false positive rates under varying imbalanced data distributions.
Finally, through qualitative analysis, we show that both LLM and human evaluators consistently rank $\name$ model's reasoning higher than other reasoning-based baselines.
\end{abstract}

\begin{links}
\link{Code}{https://github.com/martin-wey/R2Vul}
\end{links}

\section{Introduction}

Vulnerability detection (VD) is core to software security: it requires identifying insecure code constructs, reasoning about their mechanisms, and confirming the absence of flaws in safe code.
Large language models (LLMs) have demonstrated great zero-shot reasoning capabilities in language~\cite{wei2022chain, zhou2022least, kojima2022large, guo2025deepseek}, math~\cite{luo2023wizardmath, shao2024deepseekmath}, and code generation~\cite{jiang2024self, li2025structured}.
However, applying zero-shot and chain-of-thought (CoT) strategies for VD has proven disappointing, with CoT explanations often shallow or wrong~\cite{ullah2023llms, steenhoek2024comprehensive}, while detection accuracy remains low~\cite{ding2024vulnerability, nong2024chain, zhang2024prompt}.
Vulnerability detection differs from code generation as it demands a binary judgement, i.e., \emph{vulnerable vs. safe}, an inductive bias absent from generic pre‑training and instruction tuning, making additional fine-tuning essential.
Sequence classification fine-tuning (CLS) has long been applied in VD~\cite{shestov2024finetuning, chan2023transformer, lu2021codexglue}, but comes at the cost of interpretability.
The absence of any explanatory signal prevents users from trusting or debugging the prediction~\cite{doshi2017towards}.
Alternatively, recent studies have explored supervised fine-tuning (SFT) to train LLMs to generate explanations alongside predicted labels~\cite{du2024generalization, yusuf2024your, yang2024security, mao2024towards}.

Inspired by the success of reinforcement learning from AI feedback (RLAIF) in NLP~\cite{rafailov2023direct, tunstall2023zephyr} and code generation~\cite{weyssow2024codeultrafeedback, zhang2024codedpo}, we introduce $\name$, a novel approach that applies preference alignment to VD.
Instead of training solely on positive reasoning paths as SFT does, we distill a preference policy into a small student LLM by contrasting \emph{valid} and \emph{flawed} reasoning generated by a strong teacher LLM.
This additional contrastive signal explicitly teaches the student not only to generate good explanations but also to recognise and reject misleading ones, which results in higher detection accuracy than SFT.
To enable RLAIF, we contribute the first high-quality multilingual preference dataset for function-level VD by mining thousands of vulnerability-fixing commits (VFCs) from the National Vulnerability Database (NVD) across five languages: C\#, JavaScript, Java, Python, and C. 
For every vulnerable function, we prompt a teacher LLM to produce a \emph{structured reasoning} that highlights faulty code constructs, discusses the vulnerability mechanisms and their impact, and contextualizes it with CWE/CVE\footnote{CWE: Common Weaknesses Enumeration; CVE: Common Vulnerabilities and Exposures} context.
For safe functions, the teacher generates a reasoning that analyzes code safety and confirms the absence of common vulnerabilities. 
For every function, we then flip its label to generate a parallel flawed reasoning to supply contrastive reasoning pairs for RLAIF.

We validate our approach through extensive evaluation.
We fine-tune Qwen2.5-Coder-Instruct~\cite{hui2024qwen2} students at 0.5B, 1.5B, and 7B parameters and compare them with (1) four commercial static application security testing (SAST) tools, (2) CoT, CLS, and SFT applied on the same models, (3) MSIVD~\cite{du2024generalization} and VulLLM~\cite{yang2024security}, two state-of-the-art reasoning models for VD, and (4) leading proprietary LLMs (GPT-4o, GPT-4.1, Claude-4-Opus, and Claude-4-Sonnet). 
Across five languages, the 1.5B $\name$ model significantly boosts absolute macro‑F$_1$ by 5 percentage points over the strongest SFT baseline (74\% $\rightarrow$ 81\%), surpasses its 32B teacher, and even outperforms all proprietary LLMs. 
Moreover, $\name$ scales smoothly with more data: its macro‑F$_1$ rises from 73.4\% at 25\% of the training set to 81\% at full scale, while SFT plateaus at around 74\% across data regimes. 
Under extreme class imbalance, $\name$ yields far fewer false positives than SFT, and after applying a lightweight threshold calibration, it drives false positive rates down while incurring only a modest recall loss.
Finally, we show that GPT‑4o, Claude‑3.5‑Sonnet, and two security experts prefer \name{}'s explanations over CoT and MSIVD in up to 94\% of a 100‑sample qualitative set.

In summary, our contributions are the following:

\begin{itemize}
\item[--] \textbf{$\name$: A novel preference tuning approach for VD.} 
We apply RLAIF to distil a \emph{valid vs. flawed} reasoning preference signal from a teacher LLM into small student LLMs, markedly improving both detection accuracy and reasoning capabilities.

\item[--] \textbf{High-quality multilingual preference dataset.}
We release an 18k‑sample corpus spanning C\#, JavaScript, Java, Python, and C that pairs high‑confidence labels with contrastive structured reasoning to enable preference alignment for VD.

\item[--] \textbf{Comprehensive comparative study.} 
Across five languages and three model sizes (0.5B, 1.5B, and 7B), $\name$ outperforms SAST tools, CoT, CLS, SFT, MSIVD, VulLLM, and commercial LLMs; scales smoothly with data; remains robust under severe class imbalance after lightweight calibration; and produces explanations preferred by both LLM and human judges.
\end{itemize}

\section{Related Work}

\subsection{Reasoning and Instruction Tuning for VD}
\emph{Prompt‑based approaches.} 
Early work applied chain‑of‑thought (CoT) prompting to general‑purpose LLMs such as GPT‑4, but found the resulting explanations often unfaithful, resulting in limited detection accuracy~\cite{ullah2023llms,nong2024chain,ding2024vulnerability}. 
Retrieval‑augmented generation (RAG) mitigates this by injecting external security artefacts at inference time~\cite{du2024vul,wu2024effective,sun2024llm4vuln}.  
Unfortunately, none of these RAG studies release a replication package, making a rigorous comparison difficult.

\noindent \emph{Fine‑tuning approaches.} 
Yusuf and Jiang~\shortcite{yusuf2024your} applied supervised fine-tuning (SFT) but lacked structured reasoning and did not address cross-lingual generalization.
More sophisticated SFT approaches, including MSIVD~\cite{du2024generalization} and VulLLM~\cite{yang2024security}, jointly train LLMs to detect and explain vulnerabilities, while LLMVulExp~\cite{mao2024towards} focuses on explanation quality.  
However, all three prior works are confined to C/C++.

\noindent \emph{Our contribution.} 
We introduce $\name$, the first RLAIF‑based method for VD, pairing every function with a valid and a flawed structured reasoning and optimising with preference alignment. 
This teaches the model to prefer well‑founded over misleading reasoning.
$\name$ requires no retrieval at inference time, supports five programming languages, and retains strong performance even in a 0.5B‑parameter variant.  
Our experiments reveal that $\name$ outperforms strong prior approaches and models like MSIVD, VulLLM, CodeBERT, and leading commercial LLMs such as GPT-4.1 and Claude-4-Opus on all five languages.

\subsection{VD Datasets}

Label noise remains a major obstacle for training reliable VD models.
As reported by Li et al.~\shortcite{li2024cleanvul}, widely used datasets such as BigVul~\cite{fan2020ac}, CrossVul~\cite{nikitopoulos2021crossvul}, CVEFixes~\cite{cve_fixes}, and DiverseVul~\cite{chen2023diversevul} exhibit low label correctness of 25\%, 47.8\%, 51.7\%, and 60\%, respectively.
More recent datasets like SVEN~\cite{he2023large} and PrimeVul~\cite{ding2024vulnerability} achieve higher correctness of 94\% and 92\%, but are either small or restricted to C/C++.
CleanVul~\cite{li2024cleanvul} offers both scale and quality, yet omits CVE/CWE links, which are required for generating our security‑grounded structured reasoning.
We therefore curate an 18,000 samples preference dataset covering five languages with explicit CWE/CVE metadata and paired \emph{valid}–\emph{flawed} reasoning, providing the high‑confidence supervision needed for RLAIF in $\name$ with label correctness as high as 93\%.

\section{$\name$: Structured Reasoning Distillation}
\label{sec:r2vul}

\begin{figure*}
    \centering
    \includegraphics[width=.8\linewidth]{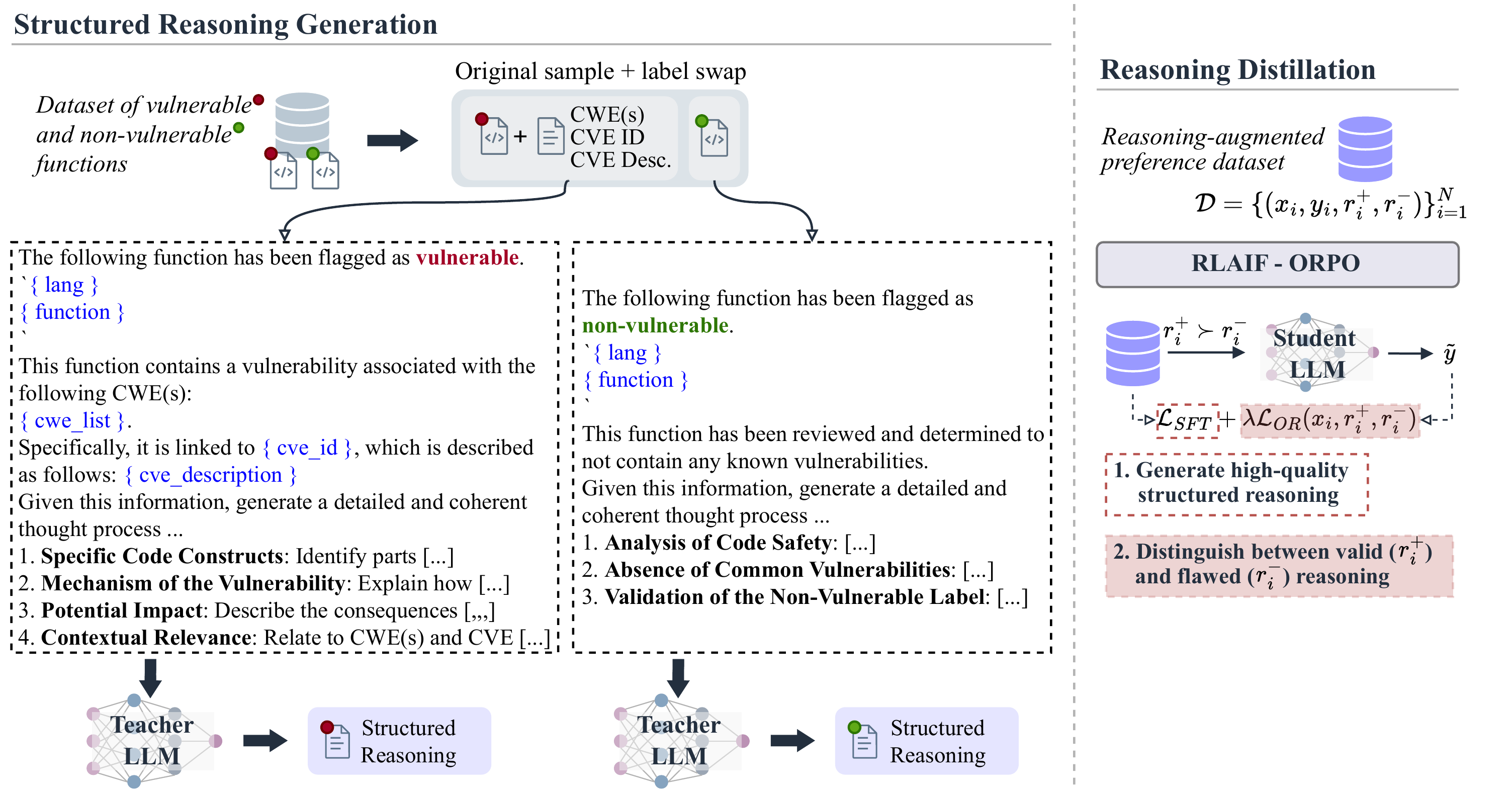}
    \caption{Overview of $\name$. A teacher LLM produces \emph{valid} reasoning $r^+_i$ using the true sample's label and \emph{flawed} reasoning $r^-_i$ via label swap. Preference tuning (RLAIF) using ORPO~\cite{hong2024orpo} to distill the reasoning-augmented dataset into a student LLM is applied.}
    \label{fig:methodology}
\end{figure*}

Our goal is to train a small code LLM that (i) accurately detects software vulnerabilities and (ii) justifies its predictions with sound, security‑aware reasoning.
We target function‑level (intraprocedural) detection and leave interprocedural reasoning to future work.
Following prior work on alignment via AI feedback~\cite{tunstall2023zephyr, weyssow2024codeultrafeedback, zhang2024codedpo}, we assume access to a larger teacher model and proceed in two stages, shown in Figure~\ref{fig:methodology}: (1) structured reasoning generation, and (2) reasoning distillation with RLAIF.

\subsection{Structured Reasoning Generation}
\label{sec:r2vul-step1}

Given a labeled dataset $\mathcal{D} = \{(x_i, y_i)\}_{i=1}^{N}$ of functions $x_i$ with vulnerability labels $y_i \in \{\mathrm{Vuln}, \mathrm{Safe}\}$, we prompt the teacher model with class‑specific templates to produce two reasoning paths per sample:
\begin{itemize}
    \item Valid reasoning $r_i^{+}$, conditioned on the true label $y_i$.
    \item Flawed reasoning $r_i^{-}$, obtained by swapping the label $y_i$.
\end{itemize}
This yields our \emph{reasoning-augmented preference dataset} $\mathcal{D} = \{(x_i, y_i, r_i^{+}, r_i^{-})\}_{i=1}^{N}$.
For example, if a function is truly vulnerable, $r_i^{-}$ asserts the code is safe and offers superficially plausible yet incorrect justifications.  
By pairing each function with both a correct and an incorrect explanation, we create direct contrastive supervision that is ideal for preference tuning.  
Because we tailor class-specific templates for the vulnerable and safe labels, the resulting reasoning paths contrast sharply in their content while following a clear, structured format. 

Figure~\ref{fig:methodology} shows excerpts of the prompt templates we use to elicit both \(r_i^{+}\) and \(r_i^{-}\) (see Appendix~\ref{sec_ap:prompts} for full prompts).
For vulnerable cases, the teacher LLM receives the function $x_i$, linked CWE list, and CVE metadata to ground its reasoning in a security-specific context rather than solely relying on code cues.
The reasoning is structured into four parts: identifying relevant code constructs, explaining the vulnerability mechanism, assessing potential impact, and linking the issue to the given CWEs and CVE identifier.
For non-vulnerable cases, the reasoning highlights safe coding practices, the absence of known issues, and justification for the non-vulnerable label.

\subsection{Reasoning Distillation with RLAIF}
\label{sec:r2vul-step2}

We fine‑tune a student LLM with RLAIF using ORPO~\cite{hong2024orpo}.
ORPO combines a standard supervised fine‑tuning (SFT) term with a single contrastive odds‑ratio loss, achieving performance superior or comparable to methods such as PPO~\cite{ziegler2019fine} or DPO~\cite{rafailov2024direct} while remaining simpler to implement and cheaper to train.

For each sample $(x_i,r_i^{+},r_i^{-})$, we construct a preference pair $(x_i;r_i^{+}\succ r_i^{-})$, where $r_i^{+}$ is the valid reasoning and $r_i^{-}$ its label-flipped counterpart.
ORPO optimises a weighted sum of an SFT loss and an odds‑ratio (OR) preference term:
\begin{equation}
\label{eq:orpo}
\mathcal{L}_{\text{ORPO}}
  = \mathcal{L}_{\text{SFT}}(x, r^{+};\theta)
  \;+\;
  \lambda\,\mathcal{L}_{\text{OR}}(x, r^{+}, r^{-};\theta),
\end{equation}
where 
\begin{equation}
\mathcal{L}_{\text{SFT}} = -\frac{1}{n} \sum_{i=1}^n \log P(r_i^+\mid x_i;\theta)
\end{equation}
\begin{equation}
\mathcal{L}_{\text{OR}} = -\frac{1}{n}\sum_{i=1}^{n} \log \sigma \left(\log\frac{\mathrm{odds}_{\theta}(r^+_i\mid x_i)}{\mathrm{odds}_{\theta}(r^-_i\mid x_i)} \right).
\end{equation}
The SFT term teaches the model to generate valid reasoning while the odds‑ratio term raises the log‑odds of selecting \(r^{+}_i\) over \(r^{-}_i\), encouraging the student to favour well‑founded explanations and suppress misleading ones.

While prior work explored preference tuning in NLP~\cite{tunstall2023zephyr, bai2022training, casper2023open} and code generation~\cite{zhang2024codedpo, weyssow2024codeultrafeedback, wei2024selfcodealign}, our work is the first to explore RLAIF with structured reasoning distillation for VD, addressing a novel domain-specific challenge.

\section{Experimental Setup}
\label{sec:exp_setup}

\subsubsection{Preference Dataset.}  
Existing VD corpora are either small‑scale~\cite{nikitopoulos2021crossvul,li2023comparison,bui2022vul4j} or heavily skewed toward C/C++~\cite{ding2024vulnerability,fan2020ac}, limiting cross‑language generalization.
We therefore curate a new multilingual preference dataset covering five languages, i.e., C\#, JavaScript, Java, Python, and C, by mining vulnerability‑fixing commits (VFCs) linked to CVEs published in the National Vulnerability Database (NVD).
Our dataset creation is guided by three key considerations.
\begin{enumerate}
    \item \textbf{Label noise.} Functions changed in a VFC are initially labeled using the common heuristic ``pre-commit=$\mathrm{Vuln}$, post-commit=$\mathrm{Safe}$'', but this heuristic results in substantial label noise given that not all functions changed in a VFC are related to a vulnerability~\cite{wang2024reposvul, li2024cleanvul}.
    We re-annotate all function pairs with GPT-4o and only retain those that the model judges directly related to the linked CVE and CWE(s).

    \item \textbf{Label quality.} To ensure label correctness, we perform a manual sanity check on 100 random samples, resulting in a 93\% label correctness, exceeding prior VD datasets such as PrimeVul (92\%)~\cite{ding2024vulnerability}, BigVul (25\%)~\cite{fan2020ac}, DiverseVul (60\%)~\cite{chen2023diversevul}, and CVEFixes (51.7\%)~\cite{cve_fixes} as reported in prior studies~\cite{li2024cleanvul,ding2024vulnerability,chen2023diversevul}.

    \item \textbf{Balanced and imbalanced splits.} To prevent overfitting on VFC paired functions, we replace fixed functions with unrelated non-vulnerable functions. 
    We opt for a balanced dataset following prior work~\cite{du2024generalization, yusuf2024your, mao2024towards} to ensure a fair evaluation of all methods.
    We stratify the corpus into 80/10/10 train/val/test splits while preserving language ratios.
    We also experiment with various class imbalance ratios to assess the robustness of our approach.
\end{enumerate}
For every function, we generate a valid and a flawed reasoning trace with Qwen2.5‑Coder‑32B‑Instruct~\cite{hui2024qwen2} as teacher model following our approach described in Section~\ref{sec:r2vul-step1}.
Overall, our corpus is the first large‑scale, multilingual dataset comprising 18,000 samples that pairs high‑confidence vulnerability labels with contrastive structured reasoning.
Full statistics and collection details appear in Appendix~\ref{sec_ap:preference_dataset}.

\subsubsection{External Test Set.} 
To further assess generalization of our approach, we create an independent Java benchmark derived from 768 manually labeled VFCs originating from various sources, e.g., NVD, Debian, Red Hat Bugzilla, from a prior study~\cite{li2023comparison}.
We first remove any commit or CVE that overlaps with our training data to avoid leakage, then extract the modified functions and keep the pre‑commit version as the candidate vulnerable sample.
Two authors with security expertise individually verify each candidate, yielding 53 high‑confidence vulnerable functions where both experts show agreement. 
We pair these with 53 non‑vulnerable Java functions randomly drawn from our own test split to form a balanced 106‑function test set. 

\subsubsection{Baselines.}
We compare $\name$ against four classes of methods widely used in recent VD research. 
\textbf{(1)} \emph{Static application security testing (SAST) tools}: we run the rule‑based industry analyzers Semgrep, SonarQube, Fortify, and Teamscale, following prior comparative studies~\cite{arusoaie2017comparison,kaur2020comparative,zhou2024comparison}.
\textbf{(2)} \emph{Sequence classification fine-tuning (CLS)}: we fine-tune the model with a linear classifier as output layer~\cite{linevul,zhou2024large}.
\textbf{(3)} \emph{Chain-of-thought (CoT)}: the model is prompted to think step-by-step~\cite{steenhoek2024comprehensive,jiang2024investigating,nong2024chain,ding2024vulnerability,zhang2024prompt}.
\textbf{(4)} \emph{Supervised fine-tuning (SFT)}: we implement SFT following prior work~\cite{du2024generalization, yusuf2024your, yang2024security, mao2024towards} by training our student LLMs on valid structured reasoning.
We also evaluate MSIVD~\cite{yang2024security} and VulLLM~\cite{du2024generalization}, two state‑of‑the‑art reasoning models, based on CodeLlama‑13B and instruction‑tuned for detection, localisation, and explanation on C/C++. 
MSIVD emits explanations at inference, whereas VulLLM outputs labels only.
We also include CodeBERT~\cite{feng2020codebert} with CLS, a strong baseline for vulnerability detection~\cite{linevul,zhou2024large,chen2023diversevul}.

\subsubsection{Teacher and Student LLMs.} 
We adopt Qwen2.5‑Coder‑32B‑Instruct~\cite{hui2024qwen2} as the teacher and use its 0.5B, 1.5B, and 7B variants as student models. 
These models achieve strong results on EvalPlus~\cite{liu2024your} and BigCodeBench~\cite{zhuo2024bigcodebench} and have shown promise for vulnerability‑related reasoning. 
We benchmark $\name$, SFT, CLS, and CoT on each student size to quantify the effect of model scaling.

\subsubsection{Metrics.} 
We use macro-F$_1$, precision, recall, false positives (FPs), and false positive rate (FPR) as main metrics.

\subsubsection{Model Calibration.} 
For experiments on model calibration, we estimate $\name$ model's confidence by comparing how likely it is to generate the first tokens of the vulnerable versus safe reasoning template.
Given the input prompt $x$, we prepend two class‑specific prefixes and compute their conditional log‑likelihoods:
\begin{center}
\small
\begin{tabular}{ll}
\textbf{Vulnerable template} &
\verb|Specific Code Constructs:| \\[2pt]
\textbf{Safe template}       &
\verb|Analysis of Code Safety:|
\end{tabular}
\end{center}
\noindent
For each template \(t\in\{\text{V},\text{S}\}\), we let
\(\ell_{t}= \tfrac{1}{|t|}\sum_{j} \log P_{\theta}(w_{j}\mid x,t_{<j})\),  
the length‑normalised log probability assigned by the model. 
We define the log‑odds margin  \(\Delta = \ell_{V}-\ell_{S}\) and convert it to a confidence score with a logistic mapping:
\[
\mathrm{conf}(x)=\sigma(\Delta)=\frac{1}{1+\exp(-\Delta)} .
\]
A function is predicted vulnerable if \(\mathrm{conf}(x)\ge\tau\) (calibration threshold).

\subsubsection{Implementation Details.} 
See Appendix~\ref{sec_ap:implementation_details}.

\section{Results and Analysis}
\label{sec:results_analysis}

\subsection{Main Results}
\label{sec:results_main}

\begin{table*}[t]
\centering
\small
\newcommand{\best}[1]{\cellcolor{green!15}\textbf{#1}} 
\begin{threeparttable}
\begin{tabular}{lcccccc}
\toprule
\textbf{Model} & \textbf{C\#} & \textbf{JavaScript} & \textbf{Java} & \textbf{Python} & \textbf{C} & \textbf{Macro-F$_1$} \\
\midrule
\multicolumn{7}{@{}l}{\textit{SAST}} \\ \addlinespace[2pt]
Semgrep & $0.0$ & $13.78$ & $3.92$ & -- & -- & $5.90$ \\
SonarQube & $0.0$ & $5.26$ & $2.57$ & -- & -- & $2.61$ \\
Fortify & $26.67$ & $19.14$ & $23.19$ & -- & -- & $23.00$ \\
Teamscale & $68.68$ & $67.52$ & $60.33$ & -- & -- & $65.51$ \\

\midrule[.5pt]
\multicolumn{7}{@{}l}{\textit{LLM baselines}} \\ \addlinespace[2pt]
VulLLM-CL-13B & $62.50$ & $59.85$ & $67.44$ & $63.96$ & $48.63$ & $60.48$ \\
MSIVD-CL-13B & $67.65$ & $66.04$ & $66.25$ & $66.09$ & $65.36$ & $66.28$ \\
CodeBERT / CLS & $76.00$ & $71.43$ & $75.60$ & $64.52$ & $70.00$ & $71.51$ \\

\addlinespace[4pt]
Claude-Sonnet-4 & $77.19$ & $64.77$ & $69.21$ & $68.78$ & $64.79$ & $68.95$ \\
Claude-Opus-4 & $70.59$ & $65.50$ & $72.39$ & $71.07$ & $66.40$ & $69.19$ \\

\addlinespace[4pt]
GPT-4o & $54.17$ & $60.00$ & $67.20$ & $64.12$ & $55.06$ & $60.31$ \\
GPT-4.1 & $63.41$ & $55.38$ & $72.00$ & $59.15$ & $58.18$ & $61.62$ \\

\midrule[.5pt]
\multicolumn{7}{@{}l}{\textit{Teacher / Student LLMs}} \\ \addlinespace[2pt]
Qwen2.5‑32B / CoT$^1$ & $74.54$ & $60.52$ & $73.83$ & $67.29$ & $56.20$ & $66.48$ \\

\addlinespace[4pt]
Qwen2.5‑0.5B / CLS & $69.77$ & $64.65$ & $78.83$ & $70.76$ & $73.45$ & $71.49$ \\
Qwen2.5‑0.5B / SFT$^1$ & $80.83$ & $71.11$ & $79.67$ & $74.14$ & $71.42$ & $75.43$ \\
\textbf{Qwen2.5‑0.5B / $\name$}$^1$ & \best{$82.80$} & \best{$\mathbf{74.60}^\dagger$} & \best{$\mathbf{85.06}^\dagger$} & \best{$74.37$} & \best{$\mathbf{74.94}^\dagger$} & \best{$\mathbf{78.35}^\dagger$} \\

\addlinespace[4pt]
Qwen2.5‑1.5B / CLS & $73.91$ & $61.54$ & $78.96$ & $72.57$ & $73.89$ & $72.17$ \\
Qwen2.5‑1.5B / SFT$^1$ & $86.28$ & $68.91$ & $81.30$ & $72.80$ & $70.73$ & $76.00$ \\
\textbf{Qwen2.5‑1.5B / $\name$}$^1$ & \best{$88.89$} & \best{$\mathbf{73.42}^\dagger$} & \best{$\mathbf{86.13}^\dagger$} & \best{$\mathbf{79.64}^\dagger$} & \best{$\mathbf{77.05}^\dagger$} & \best{$\mathbf{81.03}^\dagger$} \\

\addlinespace[4pt]
Qwen2.5‑7B / CLS & $80.85$ & $67.62$ & $81.71$ & $73.85$ & $73.85$ & $75.58$ \\
Qwen2.5‑7B / SFT$^1$ & $73.32$ & $70.58$ & $82.36$ & $75.57$ & $71.87$ & $74.74$ \\
\textbf{Qwen2.5‑7B / $\name$}$^1$ & \best{$\mathbf{86.96}^\dagger$} & \best{$\mathbf{76.90}^\dagger$} & \best{$88.05$} & \best{$\mathbf{78.26}^\dagger$} & \best{$\mathbf{77.20}^\dagger$} & \best{$\mathbf{81.47}^\dagger$} \\
\bottomrule
\end{tabular}
\begin{tablenotes}
\footnotesize
\item[$1$] Scores are averaged over three runs using identical seeds.
\item[] Qwen2.5 models refer to Qwen2.5-Coder-Instruct variants.
\end{tablenotes}
\end{threeparttable}
\caption{F$_1$ scores (\%) for five programming languages. $\dagger$ denote significant improvement of $\name$ over the corresponding SFT baseline ($p<0.05$, paired bootstrap, 10k replicas). }
\label{tab:main}
\end{table*}

We report the F$_1$ scores for five programming languages in Table~\ref{tab:main} and the macro-F$_1$.
We provide the complete Precision/Recall/F$_1$ table in Appendix~\ref{sec_ap:main}. 

\subsubsection{$\name$ vs. SAST and LLM Baselines.} 

We observe that all SAST tools perform poorly on the three tested languages. 
Teamscale scores higher than other tools, but it often reaches 100\% FPR (see Appendix~\ref{sec_ap:main}), thereby rendering it unusable in practice. 
Next, VulLLM~\cite{du2024generalization} and MSIVD~\cite{yang2024security} average only 60.48\% and 66.28\% macro-F$_1$ despite explicitly being tuned for vulnerability detection.
On the other hand, OpenAI's LLMs (GPT-4o and GPT-4.1) perform even worse, whereas Anthropic's Claude models (Claude-Sonnet-4 and Claude-Opus-4) reach higher performance. 
This finding aligns with prior work that found limited CoT reasoning capabilities in leading commercial LLMs for VD~\cite{ding2024vulnerability, nong2024chain, zhang2024prompt}.
Regardless, amongst LLM baselines, CodeBERT fine-tuned using CLS on our training dataset scores the highest (71.51\%) and surpasses all powerful commercial LLMs.

As shown in Table~\ref{tab:main}, all variants of $\name$ significantly outperform the best LLM baseline (CodeBERT) across languages, with 9.56\%, 13.31\%, and 13.93\% relative improvements in macro-F$_1$ for the 0.5B, 1.5B, and 7B models, respectively.
While fine-tuning CodeBERT with CLS already demonstrates that adapting small models to our downstream task is more beneficial than relying on advanced commercial reasoning LLMs, $\name$ further amplifies this advantage thanks to preference tuning.

\subsubsection{$\name$ vs. CLS, SFT, and Teacher LLM.} 

Our teacher LLM (Qwen2.5-32B) using CoT outperforms GPT-4o and GPT-4.1, and closely approaches Claude-Sonnet-4's performance. 
Nonetheless, all $\name$ models substantially outperform the teacher, highlighting the effectiveness of knowledge distillation and the success of our approach in specialising small models for VD. 
Across model scales, $\name$ surpasses CLS and SFT for all languages and on average.
To assess statistical robustness, we performed paired bootstrap tests between each $\name$ and SFT variants ($p < 0.05$, with 10k resamples; see Appendix~\ref{sec_ap:main} for details). 
We observe that all improvements in macro-F$_1$ achieved by $\name$ are statistically significant compared to their SFT counterparts. 
For example, the 1.5B model outperforms SFT by 5.03 absolute points in macro-F$_1$ (76.00 vs. 81.03).
Furthermore, the $\name$ 1.5B and 7B models show significant gains across four out of five languages compared to SFT.
These performance gaps underscore the benefit of incorporating an additional preference signal during fine-tuning.
Finally, we note a clear upward trend in performance with increasing model scale. 
Moreover, the narrow gaps between the 1.5B (81.03\%) and 7B (81.47\%) variants demonstrate the ability of our approach to obtain a highly effective LLM for VD even at a moderate scale.\footnote{For the remainder of our experiments, we focus on the 1.5B $\name$ variant to further study the capabilities of a small model in VD.}

\subsubsection{Generalization to External Test Set.}

\begin{table}[!t]
\centering
\newcommand{\best}[1]{\cellcolor{green!15}\textbf{#1}} 
\small
    \begin{tabular}{lrrr}
    \toprule
    \textbf{Method} & \textbf{Precision} & \textbf{Recall} & \textbf{F$_1$} \\
    \midrule
    CLS & $74.42$ & $60.38$ & $66.67$ \\
    SFT & $73.86$ & $90.57$ & $81.36$ \\
    \textbf{$\name$} & $\mathbf{80.65}^\dagger$ & $\mathbf{94.34}$ & $\mathbf{86.96}^\dagger$ \\
    \bottomrule
    \end{tabular}
\caption{Performance of Qwen2.5-1.5B on the external Java test set. $\dagger$ denotes significant improvement of $\name$ over SFT ($p<0.05$, paired bootstrap, 10k replicas).}
\label{tab:external_test_set}
\end{table}

We compare the performance of $\name$, SFT, and CLS on an external and manually-labelled Java test set in Table~\ref{tab:external_test_set}. 
This test set is complementary to ours as it originates from other data sources.
For $\name$ and SFT, we average the metrics across three runs using an identical seed and assess statistical significance using paired bootstrap tests ($p < 0.05$, 10k resamples). 
The results show that $\name$ outperforms both baselines in terms of Precision, Recall, and F$_1$, highlighting its generalization to another test set distribution.

\subsection{Data Ablation Study}
\label{sec:results_data_ablation}

\begin{figure}[!t]
    \centering
    \includegraphics[width=\linewidth]{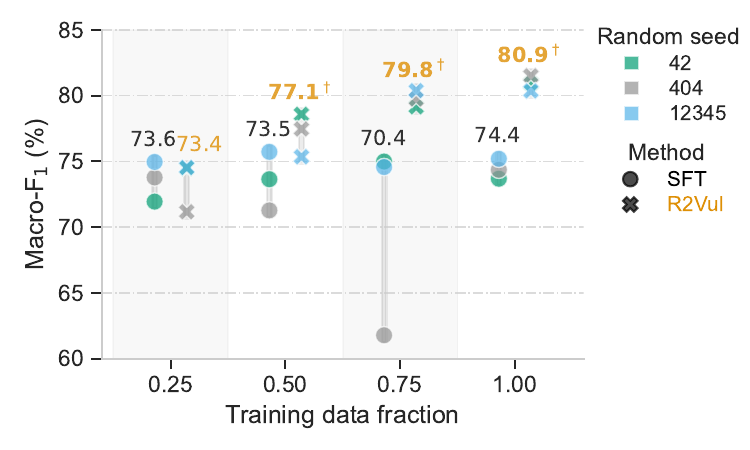}
    \caption{Macro-F$_1$ scores (\%) for Qwen2.5-1.5B fine-tuned with SFT and $\name$ at four training data fractions and three random seeds. 
    Scores in black and orange mark the seed-averaged F$_1$ for SFT and $\name$, respectively.
    $\dagger$ denotes significant improvement of $\name$ over SFT ($p<0.05$, paired bootstrap, 10k replicas).}
    \label{fig:data_ablation}
\end{figure}

To quantify data scaling behaviour, we fine-tuned Qwen2.5-1.5B on four fractions of our preference dataset (25\%, 50\%, 75\%, and 100\%) using both SFT and $\name$. 
We repeated each fine-tuning with three random seeds to mitigate variance. 
Figure~\ref{fig:data_ablation} presents the resulting macro-F$_1$ scores across data fractions and seeds for both approaches. 
Highlighted numbers refer to the seed-average F$_1$ scores (black: SFT, orange: R2Vul).

We observe that $\name$'s performance improves steadily as more training data becomes available.
Although the gains from 75\% to 100\% narrow, this trend suggests that additional data could still yield further improvements. 
These findings align with prior studies on RLHF, which reported similar data scaling trends~\cite{bai2022training, hou2024does}. 
At low data regime (25\%), SFT performs on par with $\name$, but its performance stagnates as more data is added, which suggests potential early overfitting.
In contrast, the combination of SFT loss and rewarding valid over flawed reasoning enables $\name$'s performance to scale with more data.
Across all other data regimes, $\name$ significantly outperforms SFT. 
At 100\% of the training data, SFT achieves 74.4\% macro-F$_1$, while $\name$ reaches 80.9\%.
These scores represent relative improvements in macro-F$_1$ of 1\% for SFT and 11.74\% for $\name$ compared to their respective performances at 25\% of the training data.

\subsection{Class Imbalance Impact}
\label{sec:results_class_imbalance}

\begin{figure}[!t]
    \centering
    \includegraphics[width=\linewidth]{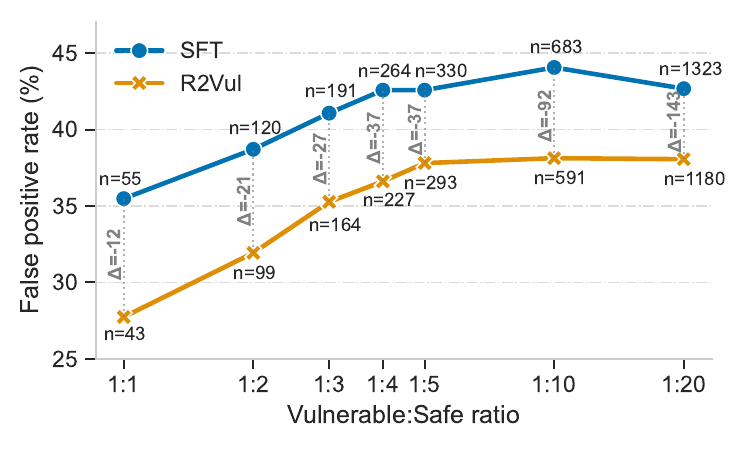}
    \caption{False positive rates (\%) of Qwen2.5-1.5B fine-tuned with SFT and R2Vul at various class imbalance ratios on our Python test set. The number of corresponding false positives are highlighted at each ratio.}
    \label{fig:class_imbalance}
\end{figure}

In real-world settings, vulnerability detection systems face highly imbalanced data distributions, where non-vulnerable functions vastly outnumber vulnerable ones. 
Moreover, prior studies have shown that high false positive rates (FPRs) are a key reason why practitioners avoid using program analysis tools~\cite{johnson2013don, imtiaz2019challenges}.
To investigate the robustness of $\name$ under such conditions, we evaluate its performance across varying levels of class imbalance. 
We focus our experiments on our Python test set and report our results in Figure~\ref{fig:class_imbalance}.
We select Python for its high popularity in open-source projects and report similar findings on Java in Appendix~\ref{sec_ap:class_imbalance}.

Evidence shows $\name$'s higher robustness compared to SFT across all class imbalance ratios. 
For instance, under balanced conditions (1:1), SFT produces 35.48\% FPs, while $\name$ yields 27.74\%, corresponding to a 28\% relative reduction. 
As the imbalance increases, FPRs for both methods plateau beyond a 1:5 ratio, with $\name$ consistently maintaining a lower rate.
Despite $\name$'s clear advantage over SFT, we observe that both methods produce a considerable number of FPs, especially under high class imbalance. 

\subsection{Model Calibration}
\label{sec:results_model_calibration}

To address the high FPRs observed under class imbalance, we apply a simple yet effective calibration technique by adjusting the decision threshold of $\name$ at inference (see Section~\ref{sec:exp_setup}).

\begin{figure}[!t]
    \centering
    \includegraphics[width=\linewidth]{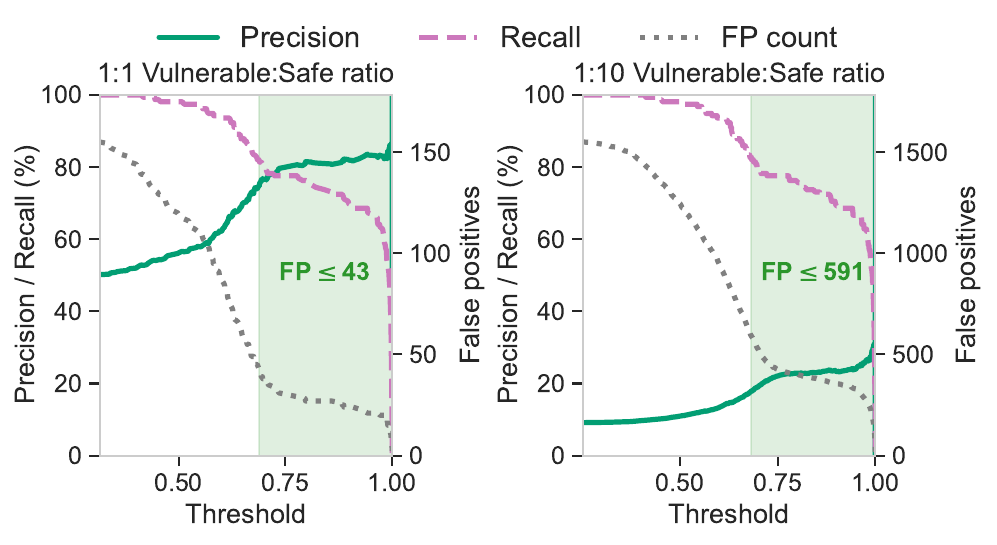}
    \caption{Effect of calibration thresholds on the precision, recall, and false positive (FP) counts of R2Vul (1.5B) on our Python test set at two class imbalance ratios (1:1 and 1:10). 
    The green areas highlight thresholds that lower FP counts compared to the raw model in Figure~\ref{fig:class_imbalance}.}
    \label{fig:calibration}
\end{figure}

Figure~\ref{fig:calibration} illustrates the effect of varying the calibration threshold on precision, recall, and FP counts for two class imbalance ratios (1:1 and 1:10) on our Python test set.
As expected, increasing the threshold reduces recall but results in markedly higher precision and significantly fewer FPs.
Notably, in both settings, we observe a wide range of thresholds (shaded in green) where the FP count falls below the raw model's FP count of 43 (for 1:1) and 591 (for 1:10), while maintaining recall above 70\%.
Within this region, precision rises steadily, reaching over 80\% in the balanced setting and over 25\% under heavy class imbalance.
While the absolute precision remains modest in the 1:10 setting due to the severity of the imbalance, this trade-off is expected and highlights how even lightweight calibration can reduce FPs meaningfully in $\name$.

\begin{table}[t]
\centering
\small
\begin{threeparttable}
\begin{tabular}{cccccc}
\toprule
\multirow{2}{*}{$\tau$} &
\multicolumn{2}{c}{False Positives} & \multicolumn{3}{c}{Metrics (\%)} \\ 
\cmidrule(lr){2-3} \cmidrule(lr){4-6}
& FP$^1$ & FPR (\%) & Precision & Recall & F$_1$ \\

\midrule
  \multicolumn{6}{@{}l}{\textit{Ratio 1:1}} \\ \addlinespace[2pt]
  -- & $43$ & $28$ & $75.29$ & $83.97$ & $79.39$ \\
  $0.73$ & $31$ & $20$ & $79.61$ & $77.56$ & $78.57$ \\
  $0.92$ & $23$ & $15$ & $83.31$ & $68.59$ & $74.83$ \\
\midrule[.5pt]
  \addlinespace[2pt]
  \multicolumn{6}{@{}l}{\textit{Ratio 1:10}} \\ \addlinespace[2pt]
  -- & $591$ & $38$ & $18.37$ & $85.26$ & $30.23$ \\
  $0.96$ & $310$ & $20$ & $24.94$ & $66.03$ & $36.20$ \\
  $0.99$ & $233$ & $15$ & $26.96$ & $55.13$ & $36.21$ \\
\bottomrule
\end{tabular}
\begin{tablenotes}
\footnotesize
\item[$1$] Out of $155$ and $1550$ benign functions for 1:1 and 1:10 imbalance ratios.
\end{tablenotes}
\end{threeparttable}
\caption{Effect of calibration thresholds ($\tau$) at two class imbalance ratios on our Python test set.}
\label{tab:calibration}
\end{table}

The above trade-off is especially important in real-world deployments.
Prior work has emphasized that FPRs must typically stay below 20\% to be acceptable to practitioners~\cite{bessey2010few, johnson2013don, christakis2016developers}.
Table~\ref{tab:calibration} explores this criterion by reporting $\name$'s performance at different calibration thresholds ($\tau$).
In the 1:1 scenario, setting $\tau{=}0.73$ lowers FPR from 28\% to 20\%, while preserving high recall (77.56\%) and F$_1$ (78.57\%). 
Under 1:10 imbalance, a threshold of $\tau{=}0.96$ achieves the same FPR target, with recall at 66.03\%.
The optimal threshold may depend on the intended use case and software system under analysis. 
In security-critical environments, prioritizing recall may be preferred to avoid overlooking true vulnerabilities.
For example, in a recent study~\cite {ami2024false}, surveyed developers reported that tolerating more FPs may be justified where the cost of a false negative is severe. 
$\name$ supports this flexibility by allowing users to adjust the threshold based on the acceptable trade-off between precision and recall.

\subsection{Qualitative Assessment of Reasoning}

\begin{table}[!t]
\centering
\small
    \begin{tabular}{lccc}
    \toprule
    & \textbf{$\name$} & \textbf{Qwen2.5-1.5B} & \textbf{MSIVD} \\
    \midrule
    \multicolumn{4}{@{}l}{\textit{Quality Criteria (GPT-4o / Claude)}} \\ \addlinespace[2pt]    
    Completeness & $4.51/4.74$ & $2.81/3.05$ & $1.24/1.09$ \\
    Clarity & $4.58/4.75$ & $3.49/3.85$ & $1.60/1.22$ \\
    Actionability & $3.84/3.81$ & $2.51/2.70$ & $1.24/1.35$ \\
    
    \midrule
    \multicolumn{4}{@{}l}{\textit{Preferences}} \\
    \addlinespace[2pt]
    GPT-4o / Claude & $94\%/93\%$ & $6\%/6\%$ & $0\%/1\%$ \\
    H1 / H2 & $92\%/93\%$ & $6\%/6\%$ & $2\%/1\%$ \\
    \bottomrule
    \end{tabular}
\caption{Reasoning quality assessment results using Claude-3.5-Sonnet, GPT-4o, and two humans (H1/H2) as evaluators.} 
\label{tab:reasoning_scores}
\end{table}

We conduct a comparative study to demonstrate that $\name$ delivers higher quality reasoning over two baselines.
We randomly select a stratified balanced set of 100 functions across all five languages to evaluate three systems: $\name$ (1.5B), Qwen2.5-1.5B with CoT, and MSIVD (CodeLlama-13B).
We use two complementary evaluation methods: \textbf{(1)} \textit{LLM‑as‑a‑Judge}: GPT‑4o and Claude‑3.5‑Sonnet score each explanation on a 0–5 scale for \textit{Completeness} (coverage of mechanism, edge cases, attack vectors), \textit{Clarity}, and \textit{Actionability} (practical guidance, e.g., pinpointed lines).
\textbf{(2)} \textit{Human ranking}: we enlisted two external security practitioners who ranked the three anonymized models' outputs to mitigate bias.

Table~\ref{tab:reasoning_scores} shows that $\name$ outperforms Qwen2.5-1.5B and MSIVD, achieving near-perfect completeness and clarity scores with Claude-3.5-Sonnet.
Both LLM judges overwhelmingly favor $\name$, ranking it first in up to 94\% of cases.
We observe nearly identical results with human evaluators (H1 / H2), who also prefer $\name$ in the majority of cases.
We discuss examples of the three models' reasoning outputs in Appendix~\ref{sec_ap:qualitative}.

\section{Conclusion and Future Work}

We introduced $\name$, the first preference tuning approach for VD.
By contrasting \emph{valid} and \emph{flawed} structured explanations generated by a teacher LLM, RLAIF with ORPO effectively distills high‑quality reasoning into student LLMs.
Our contributions also feature a new 18k‑sample high-quality multilingual preference dataset.
Across five languages, $\name$ surpasses SAST tools, CoT, CLS, SFT, recent reasoning models, and even proprietary LLMs like GPT‑4.1 and Claude-4-Opus, while a lightweight calibration step mitigates high false positives.
Human and LLM judges alike prefer $\name$'s explanations in more than 90\% of cases.
Our work targets five popular languages, leaving others (e.g., Go, Rust) and mixed‑language projects for future work.
Moreover, our study aimed at producing effective and efficient student LLMs for VD. 
Future work could investigate fine-tuning large-scale reasoning LLMs.
We hope our work can pave the way for more research on preference tuning and reasoning for vulnerability detection.

\bibliography{aaai2026}

\appendix

\section{Appendix}

\subsection{Preference Dataset}
\label{sec_ap:preference_dataset}

\subsubsection{Data Collection.} 
We mine NVD entries containing a corresponding \emph{patch} (or VFC: vulnerability-fixing commit) on GitHub. 
We target five programming languages widely used in open-source projects: C\#, JavaScript, Java, Python, and C.
Table~\ref{tab:raw_dataset_statistics} presents the raw dataset statistics, which features a broad range of CVEs and CWEs.
Following prior work~\cite{chen2023diversevul}, we first labelled functions by considering pre-commit functions modified in a VFC as vulnerable (V) and all others as non-vulnerable (S).

To mitigate the low label correctness of this heuristic-based approach, we leveraged GPT-4o for re-annotation. 
The goal is to mitigate label noise and ensure that only functions directly responsible for vulnerabilities in VFCs are labeled as vulnerable.
First, we apply processing to our raw dataset by filtering out test-related functions, deduplicating samples via MD5 hash, and removing functions 4,096 tokens to avoid training on unreasonably long functions. 
Then, each function pair is labeled using GPT-4o based on its pre/post-commit versions, CWE/CVE identifiers, and CVE description linked to the VFC. 
The model assigns a vulnerability score from 0 (unrelated) to 4 (directly responsible), allowing us to filter out noisy labels and retain high-confidence vulnerable samples (score = 4).
The full prompt appears in Table~\ref{tab:apx_prompt_labelling}.
Table~\ref{tab:cleanvul_nvd_statistics} presents the number of vulnerable functions under different score thresholds.
We set a conservative threshold of $\tau = 4$ to select vulnerable functions and ensure high label correctness.

To construct our final dataset, we merged the validated vulnerable functions with non-vulnerable functions from the original dataset (Table~\ref{tab:raw_dataset_statistics}) to form a balanced dataset.
This prevents overfitting to function pairs during fine-tuning by including non-vulnerable examples unrelated to the post-commit version of the vulnerability.
We opted for a balanced dataset following prior work~\cite{du2024generalization, yusuf2024your, mao2024towards} to ensure a fair evaluation of model performance across vulnerable and non-vulnerable functions. 
We randomly splitted the dataset into train/validation/test sets using an 80:10:10 ratio, preserving the distribution of samples per programming language.
Finally, we generated valid and flawed reasoning for all samples using Qwen2.5-Coder-32B-Instruct following our approach described in Section~\ref{sec:r2vul}.

\subsubsection{Annotation Sanity Check.} 
To assess the reliability of LLM-generated labels, we conducted a sanity check with three authors with expertise in software security.
We randomly sampled 100 vulnerable functions (20 per language) and asked annotators to accept, reject, or mark labels as uncertain.
The results were as follows: 89/4/7, 86/13/1, and 93/3/4 (Accept/Uncertain/Reject).
Using a majority vote, we found that 93\% of labels were confirmed correct, while 1\% were rejected, and 6\% were uncertain.
These findings confirm the high accuracy of LLM-generated labels, aligning with prior work~\cite{li2024cleanvul}, which reports up to 97.5\% correctness when using LLMs for automated annotation.
Furthermore, our dataset demonstrates higher correctness than existing SVD datasets, including PrimeVul (92\%)~\cite{ding2024vulnerability}, BigVul (25\%)~\cite{fan2020ac}, DiverseVul (60\%)~\cite{chen2023diversevul}, and CVEFixes (51.7\%)~\cite{cve_fixes} as reported in prior studies~\cite{li2024cleanvul,ding2024vulnerability,chen2023diversevul}.

\begin{table}[!t]
\centering
\small
\begin{tabular}{lrrrr}
\toprule
Language & $\tau=1$ & $\tau=2$ & $\tau=3$ & $\tau=4$ \\
\midrule
C & 6,715 & 6,124 & 5,425 & \textbf{4,979} \\
Python & 2,873 & 2,328 & 1,762 & \textbf{1,552} \\
JavaScript & 2,834 & 2,254 & 1,462 & \textbf{1,093} \\
Java & 2,736 & 2,321 & 1,839 & \textbf{1,540} \\
C\# & 337 & 308 & 264 & \textbf{223} \\
\midrule
\end{tabular}
\caption{Number of vulnerable functions at different score thresholds ($\tau$) after GPT-4o re-labeling.} 
\label{tab:cleanvul_nvd_statistics}
\end{table}

\begin{table}[!t]
\centering
\small
\begin{tabular}{lrrrrr}
\toprule
Language & \# V & \# S & CVE & CWE & \# Pairs \\
\midrule
C & 12,670 & 215,675 & 3,719 & 162 & 7,800 \\
Python & 8,127 & 81,465 & 1,008 & 179 & 5,397 \\
JavaScript & 20,076 & 71,566 & 947 & 133 & 4,187 \\
Java & 5,325 & 59,524 & 697 & 139 & 3,958 \\
C\# & 656 & 5,240 & 93 & 46 & 391 \\
\midrule
\end{tabular}
\caption{Raw NVD dataset statistics (\# V/S: number of vulnerable/safe functions, CVE/CWE: number of unique CVE/CWE).} 
\label{tab:raw_dataset_statistics}
\end{table}

\subsection{Implementation Details.} 
\label{sec_ap:implementation_details}

We use HuggingFace TRL (0.12.1)~\cite{vonwerra2022trl} and TGI (3.3.4) for fine-tuning and inference with CLS, SFT, and $\name$ (using ORPO~\cite{hong2024orpo}).
We use LoRA~\cite{hu2021lora} for efficient fine-tuning with $r=8$, $\alpha=16$, and a dropout of $0.05$. 
We use a linear scheduler without warmup, setting the learning rate to 5e-5 for CLS and 3e-4 for SFT/ORPO.
We train using a batch size of 16 for CLS, and 2 for SFT/ORPO, with a maximum of 10 (CLS) and 5 epochs (SFT/ORPO).
For experiments with ORPO, we set $\lambda=0.3$ following ORPO's paper recommendation.
For more details, see Appendix~\ref{sec_ap:hp}, where we report a hyperparameter sensitivity analysis of ORPO.
For CodeBERT, we set the maximum input sequence length to 512. 
For other LLMs, we limit the input function to 4,096 tokens and the total input prompt to 32,768 tokens.
For inference, we use nucleus sampling with a temperature of 0.2 and a maximum generation length of 2,048 tokens.

All experiments were conducted on a DGX machine running Ubuntu 22.04.4, with an Intel(R) Xeon(R) Platinum 8480C CPU and 2TB of RAM. 
We used a single NVIDIA H100 GPU with 80GB of memory for all experiments.
We share our datasets, model checkpoints, and code in our replication package: \url{https://github.com/martin-wey/R2Vul}.

\subsection{Main Results Details}
\label{sec_ap:main}

We report precision, recall, and F$_1$ scores for our main experiment in Table~\ref{tab:apx_main}. 
We compute paired bootstrap with 10,000 replicas for each programming language and on macro-averaged metrics between each $\name$ and SFT LLM. 
For each model, we run inference 3 times using an identical seed and concatenate the labels/predictions from each run before computing the paired bootstrap test. 
We performed similar paired bootstrap tests across experiments and report statistical significance with $p < 0.05$. 

\begin{sidewaystable*}[!t]
\centering
\large
\newcommand{\best}[1]{\cellcolor{green!15}\textbf{#1}} 
\newcommand{\second}[1]{\underline{#1}} 
\resizebox{\textwidth}{!}{%
\begin{tabular}{lllllllllllllllllll}
\toprule
 & \multicolumn{3}{c}{\textbf{C\#}} & \multicolumn{3}{c}{\textbf{JS}} & \multicolumn{3}{c}{\textbf{Java}} & \multicolumn{3}{c}{\textbf{Py}} & \multicolumn{3}{c}{\textbf{C}} & \multicolumn{3}{c}{\textbf{Macro Avg}} \\

\textbf{Model} & \multicolumn{1}{c}{Precision} & \multicolumn{1}{c}{Recall} & \multicolumn{1}{c}{F$_1$} & \multicolumn{1}{c}{Precision} & \multicolumn{1}{c}{Recall} & \multicolumn{1}{c}{F$_1$} & \multicolumn{1}{c}{Precision} & \multicolumn{1}{c}{Recall} & \multicolumn{1}{c}{F$_1$} & \multicolumn{1}{c}{Precision} & \multicolumn{1}{c}{Recall} & \multicolumn{1}{c}{F$_1$} & \multicolumn{1}{c}{Precision} & \multicolumn{1}{c}{Recall} & \multicolumn{1}{c}{F$_1$} & \multicolumn{1}{c}{Precision} & \multicolumn{1}{c}{Recall} & \multicolumn{1}{c}{F$_1$} \\
\midrule
\multicolumn{19}{@{}l}{\textit{SAST}} \\ \addlinespace[2pt]
Semgrep & $0.0$ & $0.0$ & $0.0$ & $100.0$ & $0.74$ & $13.78$ & $100.0$ & $2.0$ & $3.92$ & -- & -- & -- & -- & -- &  -- & $66.67$ & $0.91$ & $5.90$ \\
SonarQube & $0.0$ & $0.0$ & $0.0$ & $100.0$ & $0.27$ & $5.26$ & $100.0$ & $1.30$ & $2.57$ & -- & -- & -- & -- & -- & -- & $66.67$ & $0.79$ & $2.61$ \\
Fortify & $57.10$ & $17.40$ & $26.67$ & $41.90$  & $12.40$ & $19.14$ & $51.10$ & $15.0$ & $23.19$ & -- & -- & -- & -- & -- & -- & $50.03$ & $14.93$ & $23.00$ \\
Teamscale & $52.30$ & $100.0$ & $68.68$ & $51.20$ & $99.10$ & $67.52$ & $54.80$ & $67.10$ & $60.33$ & -- & -- & -- & -- & -- & -- & $52.77$ & $88.73$ & $65.51$ \\
\addlinespace[2pt]

\midrule[.5pt]
\multicolumn{19}{@{}l}{\textit{LLMs}} \\ \addlinespace[2pt]

CodeBERT & $70.37$ & $82.61$ & $76.00$ & $71.43$ & $71.43$ & $71.43$ & $79.71$ & $71.90$ & $75.60$ & $73.17$ & $57.69$ & $64.52$ & $74.94$ & $65.67$ & $70.00$ & $73.92$ & $69.86$ & $71.51$ \\
VulLLM-CL-13B & $60.00$ & $65.22$ & $62.50$ & $48.52$ & $78.10$ & $59.85$ & $60.31$ & $76.47$ & $67.44$ & $52.94$ & $80.77$ & $63.96$ & $48.02$ & $49.25$ & $48.63$ & $53.96$ & $69.96$ & $60.48$ \\
MSIVD-CL-13B & $51.11$ & $100.00$ & $67.65$ & $49.30$ & $100.00$ & $66.04$ & $49.84$ & $100.00$ & $66.25$ & $49.83$ & $98.08$ & $66.09$ & $48.65$ & $99.57$ & $65.36$ & $50.55$ & $99.53$ & $66.28$ \\

\addlinespace[4pt]

Claude-Sonnet-4 & $64.71$ & $95.65$ & $77.19$ & $51.70$ & $86.67$ & $64.77$ & $54.51$ & $94.77$ & $69.21$ & $55.51$ & $90.38$ & $68.78$ & $50.35$ & $90.83$ & $64.79$ & $55.36$ & $81.66$ & $68.95$ \\
Claude-Opus-4 & $64.29$ & $78.26$ & $70.59$ & $60.48$ & $71.43$ & $65.50$ & $61.36$ & $88.24$ & $72.39$ & $62.32$ & $82.69$ & $71.07$ & $53.29$ & $88.06$ & $66.40$ & $60.35$ & $81.74$ & $61.62$ \\

\addlinespace[4pt]

GPT-4o & $52.00$ & $56.52$ & $54.17$ & $51.72$ & $71.43$ & $60.00$ & $57.08$ & $81.70$ & $67.20$ & $59.24$ & $69.87$ & $64.12$ & $50.63$ & $60.34$ & $55.06$ & $54.62$ & $67.97$ & $60.31$ \\
GPT-4.1 & $72.22$ & $56.52$ & $63.41$ & $60.00$ & $51.43$ & $55.38$ & $73.47$ & $70.59$ & $72.00$ & $65.63$ & $53.85$ & $59.15$ & $53.80$ & $63.33$ & $58.18$ & $65.02$ & $59.14$ & $61.62$ \\
\addlinespace[2pt]

\midrule[.5pt]
\multicolumn{19}{@{}l}{\textit{Teacher / Student LLMs}} \\ \addlinespace[2pt]

Qwen2.5-32B / CoT & $68.10\pm4.40$ & $82.61\pm4.35$ & $74.54\pm2.64$ & $51.69\pm0.37$ & $73.02\pm1.10$ & $60.52\pm0.19$ & $67.51\pm.11$ & $81.48\pm2.64$ & $73.83\pm1.07$ & $60.48\pm0.70$ & $75.85\pm2.06$ & $67.29\pm0.96$ & $56.51\pm0.55$ & $55.93\pm2.34$ & $56.20\pm1.08$ & $60.86\pm1.19$ & $73.78\pm1.84$ & $66.48\pm0.63$ \\

\addlinespace[4pt]

Qwen2.5-0.5B / CoT & $0.0$ & $0.0$ & $0.0$ & $0.0$ & $0.0$ & $0.0$ & $0.0$ & $0.0$ & $0.0$ & $0.0$ & $0.0$ & $0.0$ & $0.0$ & $0.0$ & $0.0$ & $0.0$ & $0.0$ & $0.0$ \\
Qwen2.5-0.5B / CLS & $75.00$ & $65.22$ & $69.77$ & $68.82$ & $60.95$ & $64.65$ & $78.57$ & $79.08$ & $78.83$ & $65.05$ & $77.56$ & $70.76$ & \best{$72.31$} & $74.63$ & $73.45$ & $71.95$ & $71.49$ & $71.49$ \\
Qwen2.5-0.5B / SFT & $76.69\pm3.02$ & \best{$85.51\pm2.51$} & $80.83\pm2.20$ & $61.26\pm1.09$ & \best{$84.76\pm0.95$} & $71.11\pm0.47$ & $72.20\pm0.76$ & \best{$88.89\pm3.27$} & $79.67\pm1.68$ & $70.82\pm1.08$ & \best{$77.78\pm0.98$} & $74.14\pm1.04$ & $62.46\pm0.37$ & \best{$83.37\pm0.56$} & $71.42\pm0.42$ & $68.69\pm0.45$ & \best{$84.06\pm0.23$} & $75.43\pm0.27$ \\
\textbf{Qwen2.5-0.5B / $\name$} & \best{$89.82\pm0.30$} & $76.81\pm2.51$ & \best{$82.80\pm1.60$} & \best{$71.96\pm0.97$} & $77.46\pm2.75$ & \best{$74.60\pm1.78^\dagger$} & \best{$81.73\pm0.36$} & $88.67\pm0.75$ & \best{$85.06\pm0.51^\dagger$} & \best{$73.14\pm0.39$} & $75.64\pm0.64$ & \best{$74.37\pm0.21$} & $71.14\pm0.16$ & $79.18\pm0.44$ & \best{$74.94\pm0.28^\dagger$} & \best{$77.56\pm0.41$} & $79.55\pm1.25$ & \best{$78.35\pm0.83^\dagger$} \\

\addlinespace[4pt]

Qwen2.5-1.5B / CoT & $55.56$ & $43.48$ & $48.78$ & $53.00$ & $59.05$ & $55.86$ & $59.17$ & $65.36$ & $62.11$ & $57.67$ & $69.87$ & $63.19$ & $53.25$ & $75.05$ & $62.30$ & $55.93$ & $62.56$ & $58.45$ \\
Qwen2.5-1.5B / CLS & $73.91$ & $73.91$ & $73.91$ & $66.67$ & $57.14$ & $61.54$ & $78.20$ & $79.74$ & $78.96$ & $67.21$ & $78.85$ & $72.57$ & \best{$73.57$} & $74.20$ & $73.89$ & $71.92$ & $72.77$ & $72.17$ \\
Qwen2.5-1.5B / SFT & $81.83\pm2.02$ & \best{$91.30\pm4.35$} & $86.28\pm2.56$ & \best{$69.58\pm0.34$} & $68.25\pm0.55$ & $68.91\pm0.32$ & $74.46\pm1.30$ & $89.54\pm1.73$ & $81.30\pm1.37$ & $67.15\pm1.04$ & $79.49\pm1.28$ & $72.80\pm1.11$ & $65.64\pm1.04$ & $76.69\pm0.89$ & $70.73\pm0.97$ & $71.73\pm0.54$ & $81.06\pm0.66$ & $76.00\pm0.08$ \\
\textbf{Qwen2.5-1.5B / $\name$} & \best{$90.91\pm0.0$} & $86.96\pm0.0$ & \best{$88.89\pm0.0$} & $69.29\pm0.54$ & \best{$78.10\pm2.52$} & \best{$73.42\pm1.39^\dagger$} & \best{$80.52\pm1.78$} & \best{$92.56\pm0.38$} & \best{$86.13\pm1.09^\dagger$} & \best{$75.38\pm0.50$} & \best{$84.40\pm0.37$} & \best{$79.64\pm0.44^\dagger$} & $70.17\pm0.26$ & \best{$85.43\pm0.49$} & \best{$77.05\pm0.09^\dagger$} & \best{$77.25\pm0.42$} & \best{$85.50\pm0.57$} & \best{$81.03\pm0.47^\dagger$} \\

\addlinespace[4pt]

Qwen2.5-7B / CoT & $63.33$ & $82.61$ & $71.70$ & $54.54$ & $57.14$ & $55.81$ & $59.00$ & $77.12$ & $66.86$ & $59.28$ & $73.72$ & $65.71$ & $53.18$ & $65.88$ & $58.86$ & $57.87$ & $71.29$ & $63.79$ \\
Qwen2.5-7B / CLS & $79.17$ & $82.61$ & $80.85$ & $67.62$ & $67.62$ & $67.62$ & $76.57$ & $87.58$ & $81.71$ & $71.01$ & $76.92$ & $73.85$ & \best{$72.48$} & $75.27$ & $73.85$ & $73.37$ & $78.00$ & $75.58$ \\
Qwen2.5-7B / SFT & $77.58\pm6.53$ & $69.57\pm4.35$ & $73.32\pm5.03$ & \best{$77.56\pm0.98$} & $64.76\pm1.90$ & $70.58\pm1.53$ & $81.31\pm0.44$ & $83.44\pm1.36$ & $82.36\pm0.84$ & \best{$76.14\pm0.59$} & $75.00\pm0.64$ & $75.57\pm0.49$ & $72.11\pm0.73$ & $71.64\pm0.77$ & $71.87\pm0.58$ & $76.94\pm1.50$ & $72.88\pm0.95$ & $74.74\pm1.17$ \\
\textbf{Qwen2.5-7B / $\name$} & \best{$86.96\pm0.0$} & \best{$86.96\pm0.0$} & \best{$86.96\pm0.0^\dagger$} & $72.47\pm0.35$ & \best{$81.90\pm0.0$} & \best{$76.90\pm0.20^\dagger$} & \best{$82.88\pm0.31$} & \best{$93.90\pm0.38$} & \best{$88.05\pm0.31$} & $74.28\pm0.66$ & \best{$82.69\pm0.0$} & \best{$78.26\pm0.36^\dagger$} & $71.61\pm0.50$ & \best{$83.72\pm0.49$} & \best{$77.20\pm0.50^\dagger$} & \best{$77.64\pm0.22$} & \best{$85.84\pm0.09$} & \best{$81.47\pm0.16^\dagger$} \\
\bottomrule
\end{tabular}
}
\caption{Precision, recall, and F$_1$ scores (\%) for five programming languages. $\dagger$ denote significant improvement of $\name$ over the corresponding SFT baseline ($p<0.05$, paired bootstrap, 10k replicas). }
\label{tab:apx_main}
\end{sidewaystable*}

\subsection{Hyperparameter Sensitivity Analysis}
\label{sec_ap:hp}

We analyse $\name$'s training dynamics across epochs for different $\lambda$ values ($\lambda \in {0.1, \dots, 1.0}$).
The $\lambda$ hyperparameter controls the balance between SFT and preference tuning in ORPO, and we vary it to explore low to high contributions of preference tuning.

\begin{figure}[!t]
    \centering
    \includegraphics[width=\linewidth]{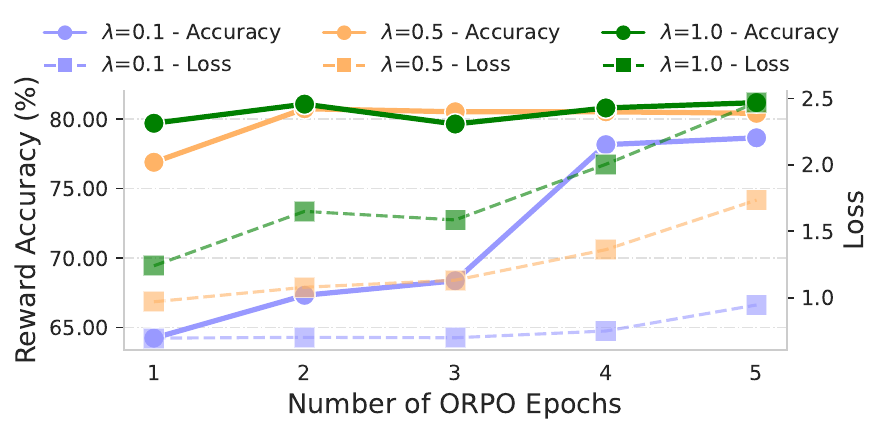}
    \caption{Evolution of reward accuracy and loss using ORPO on the validation set across epochs for $\lambda \in \{0.1, 0.5, 1.0\}$.}
    \label{fig:apx_orpo_loss_reward}
\end{figure}

Figure~\ref{fig:apx_orpo_loss_reward} shows how reward accuracy and validation loss evolve during training for Qwen2.5-1.5B with $\lambda \in {0.1, 0.5, 1.0}$.
Across all settings, the model starts to overfit after three epochs, as seen by rising validation loss.
This overfitting correlates with a rise in reward accuracy, which suggests that the model becomes increasingly effective at distinguishing between valid and flawed reasoning.
Notably, the small gap between $\lambda=0.5$ and $\lambda=1.0$ implies that pushing $\lambda$ too high may not offer any additional benefit.

\begin{figure}[!t]
    \centering
    \includegraphics[width=\linewidth]{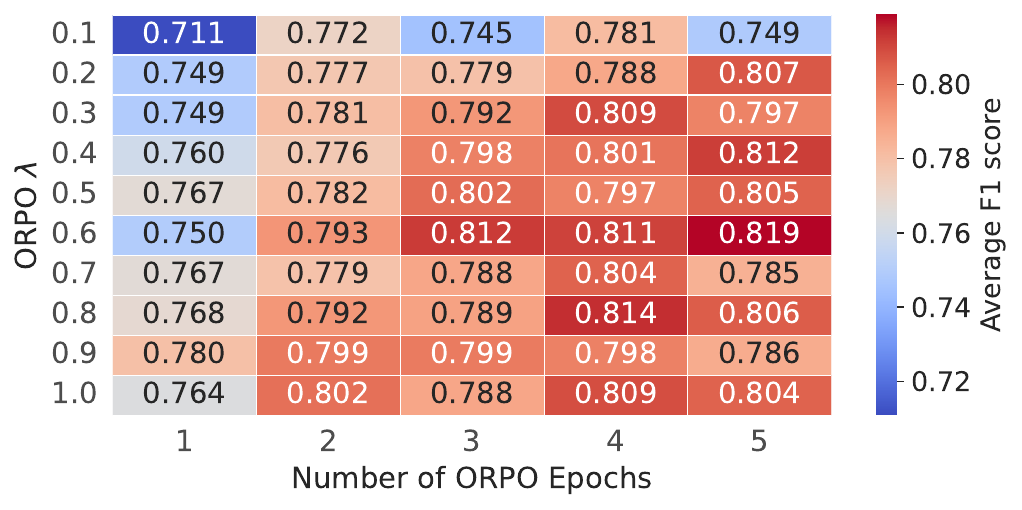}
    \caption{Evolution of the macro-F$_1$ score on the test set with various ORPO $\lambda$ values and number of training epochs.}
    \label{fig:apx_orpo_heatmap}
\end{figure}

Figure~\ref{fig:apx_orpo_heatmap} further shows that peak macro-F$_1$ scores on the test set are consistently reached after four to five epochs across all $\lambda$ values.
The highest macro-F$_1$ score is achieved at $\lambda=0.6$ after five epochs, while both lower and higher $\lambda$ values yield weaker results.
Thus, although R2Vul performed well in our experiments (with $\lambda=0.3$), its performance could improve with further hyperparameter tuning.
Moreover, this suggests that the recommended $\lambda \in \{0.1, 0.2, 0.3\}$ from ORPO's seminal paper~\cite{hong2024orpo} is not optimal for our task.

\subsection{Additional Class Imbalance Experiments}
\label{sec_ap:class_imbalance}

\begin{figure}
    \centering
    \includegraphics[width=\linewidth]{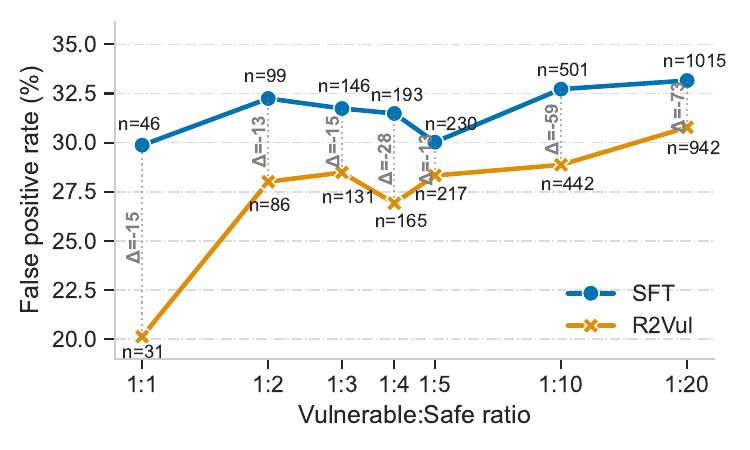}
    \caption{False positive rates (\%) of Qwen2.5-1.5B fine-tuned with SFT and R2Vul at various class imbalance ratios on our Java test set. The numbers of corresponding false positives are highlighted at each ratio.}
    \label{fig:apx_class_imbalance_java}
\end{figure}

We report additional results for our class imbalance experiments on Java in Figure~\ref{fig:apx_class_imbalance_java}.
We observe similar results to those obtained on our Python test set, with $\name$ outperforming SFT in all class imbalance ratios.

\subsection{Qualitative Analysis Examples}
\label{sec_ap:qualitative}

\begin{figure*}[!t]
    \centering
    \includegraphics[width=0.75\linewidth]{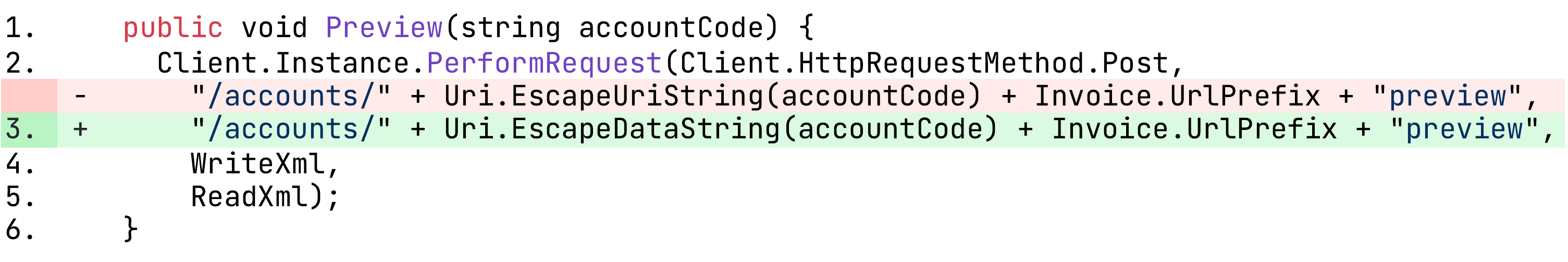}
    \caption{Vulnerable C\# function associated with CVE-2017-0907 and CWE-918 (Server-Side Request Forgery (SSRF)).}
    \label{fig:example_c_sharp}
    
    \vspace{1em}
    
    \small
    \begin{tabular}{p{\textwidth}}
    \hline
    \multicolumn{1}{c}{\cellcolor{blue!10} \textbf{\small R2Vul 1.5B (ours)}} \\
    \begin{minipage}{\textwidth}
    \vspace{.2em}

\verb|<thinking>|\\
1. **Specific Code Constructs**: The specific code construct contributing to the vulnerability is the use of `Uri.EscapeUriString(accountCode)` in the URL construction. This function is intended to escape data for inclusion in a URI component, but it does not provide adequate protection against certain types of injection attacks.\\

2. **Mechanism of the Vulnerability**: The mechanism of the vulnerability lies in the incorrect use of `Uri.EscapeUriString`. This function only escapes characters that have special meanings in URIs, such as `?', `/`, `\&`, etc., but it does not prevent other potentially harmful characters from being included in the URL if they are not properly handled elsewhere in the application. In the context of the provided CVE, this could mean that another part of the application uses `accountCode` without further encoding or validation, leading to potential injection of malicious content into the URL.\\

3. **Potential Impact**: Exploiting this vulnerability could allow an attacker to inject arbitrary parameters into the URL used for requesting invoice previews. This could lead to unauthorized access to sensitive information, such as financial records, or manipulation of the request to perform actions on behalf of the server (e.g., downloading files). Depending on the server's configuration, this could result in full control over the server or leakage of sensitive data.\\

4. **Contextual Relevance**: This vulnerability aligns with CWE-918, which describes Improper Neutralization of Special Elements used in an OS Command ('OS Command Injection'). However, in this case, it more closely relates to improper handling of URLs, specifically the misuse of `Uri.EscapeUriString`. The CVE-2017-0907 description highlights the issue in the Recurly Client .NET Library where the library incorrectly handles certain inputs, allowing attackers to craft requests that can lead to unauthorized access or manipulation. The provided code snippet demonstrates a similar pattern of using `Uri.EscapeUriString` without ensuring that all parts of the URL are properly encoded, which is consistent with the described vulnerability.\\
\verb|</thinking>|\\
\verb|<output>|
YES
\verb|</output>|

    \end{minipage} 
    \\
    \multicolumn{1}{c}{\cellcolor{cyan!10} \textbf{\small Qwen2.5-Coder-1.5B-Instruct (CoT+Reflection)}} \\
    \begin{minipage}{\textwidth}
    \vspace{.2em}

\verb|<Thinking>|\\
The given function `Preview` performs an HTTP POST request to a specific endpoint using the `Client.Instance.PerformRequest` method. The endpoint URL includes the escaped account code and a prefix followed by "preview". There is no indication of any potential security issues such as SQL injection, cross-site scripting (XSS), or other common web vulnerabilities.\\
However, the function does not contain any explicit checks or validations for the input `accountCode`. It simply escapes the URI string before appending it to the endpoint URL. While this can prevent some types of attacks, it does not guarantee that the input is safe from all possible attacks.\\
Therefore, based on the information provided, we cannot definitively say that the function is vulnerable without additional context or further analysis.\\
\verb|</thinking>|\\
\verb|<reflection>|\\
The function does not contain any direct evidence of a vulnerability. However, it lacks proper validation of the input `accountCode`, which could potentially lead to security risks if the input is not properly sanitized or validated.\\
\verb|</reflection>|\\
\verb|<output>|
NO
\verb|</output>|

    \end{minipage} 
    \\
    \multicolumn{1}{c}{\cellcolor{cyan!10} \textbf{\small MSIVD (CodeLlama-13B)}} \\
    \begin{minipage}{\textwidth}
    \vspace{.2em}
Yes.
\verb|</s>|\\
The vulnerability is type CWE-399, complexity Medium, with the following description: '''The ALSA control implementation in the ALSA subsystem in the ALSA subsystem in the ALSA subsystem in the ALSA subsystem \texttt{...}

    \vspace{.2em}
    \end{minipage} 
    \\
    \hline
    \end{tabular}
    \captionof{table}{Comparison of reasoning output of $\name$ (1.5B), Qwen2.5-Coder-1.5B-Instruct, and MSIVD (CodeLlama-13B) for the vulnerable C\# function above.}
    \label{tab:example_c_sharp}
    
\end{figure*}

\begin{figure*}[!t]
    \centering
    \includegraphics[width=0.6\linewidth]{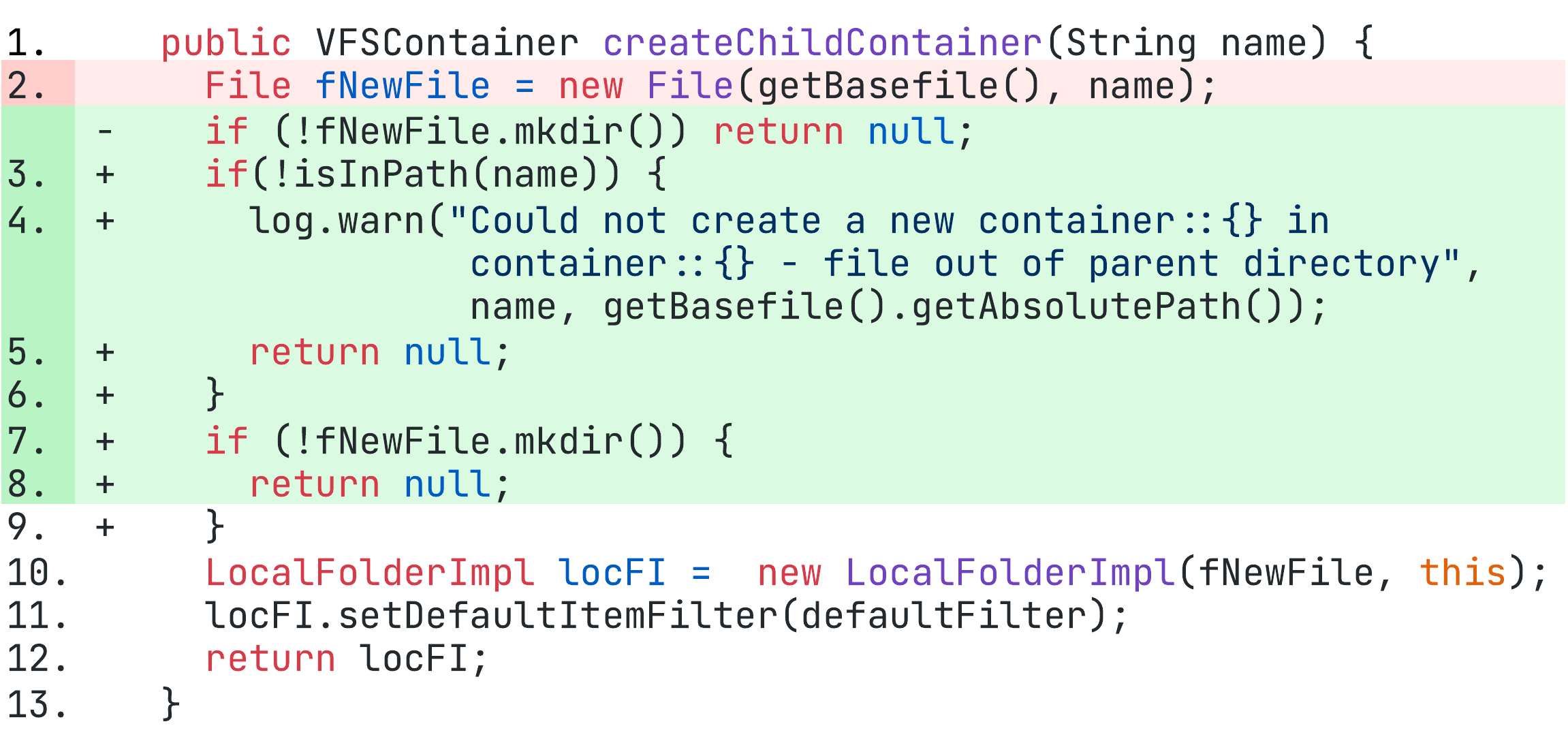}
    \caption{Vulnerable Java function associated with CVE-2021-41242 and CWE-22 (Improper Limitation of a Pathname to a Restricted Directory (Path Traversal)).}
    \label{fig:example_java}
    
    \vspace{1em}
    
    \small
    \begin{tabular}{p{\textwidth}}
    \hline
    \multicolumn{1}{c}{\cellcolor{blue!10} \textbf{\small R2Vul 1.5B (ours)}} \\
    \begin{minipage}{\textwidth}
    \vspace{.2em}
\verb|<thinking>|\\
1. **Specific Code Constructs**: The primary issue lies in the line `File fNewFile = new File(getBasefile(), name);`. Here, the `name` parameter is directly used without any form of validation or sanitization to construct a file path. Additionally, the method `mkdir()` is called on `fNewFile`, which attempts to create a directory at the constructed path if it does not exist.\\
\\
2. **Mechanism of the Vulnerability**: The vulnerability arises from the lack of input validation for the `name` parameter. An attacker can exploit this by providing a specially crafted string that includes directory traversal sequences (such as `../`). When `new File(getBasefile(), name)` constructs the file path using such a malicious `name`, it can point to any location on the filesystem accessible by the application. The subsequent call to `mkdir()` then creates or attempts to create the directory specified by the attacker, potentially overwriting existing files or accessing sensitive data.\\
\\
3. **Potential Impact**: Exploiting this vulnerability could allow an attacker to perform actions such as deleting critical system files, modifying configuration files, or even accessing sensitive data stored on the server's filesystem. In the context of OpenOLAT, as described in CVE-2021-39180, attackers could gain unauthorized access to course structures, student data, or other resources, leading to potential data breaches and service disruptions.\\
\\
4. **Contextual Relevance**: This vulnerability aligns with CWE-22, which describes "Improper Limitation of a Pathname to a Restricted Directory ('Path Traversal')". The CWE highlights the risk of constructing file paths using unvalidated user input, which can lead to unauthorized access to files outside the intended directory. The specific CVE-2021-39180 further contextualizes this issue within the OpenOLAT application, detailing how an attacker could exploit this flaw to manipulate file operations and potentially gain administrative privileges or access to sensitive information.\\
\verb|</thinking>|\\
\verb|<output>|
YES
\verb|</output>|
    \end{minipage} 
    \\
    \multicolumn{1}{c}{\cellcolor{cyan!10} \textbf{\small Qwen2.5-Coder-1.5B-Instruct (CoT+Reflection)}} \\
    \begin{minipage}{\textwidth}
    \vspace{.2em}
\verb|<Thinking>|\\
The given Java method `createChildContainer` creates a new directory named `name' under the base file specified by `getBasefile()`. It uses the `mkdir()` method to attempt to create the directory. If the directory creation fails, it returns `null`.\\
However, the method does not handle exceptions properly. If the directory creation fails due to permissions issues or other reasons, it will simply return `null`, which might lead to unexpected behavior or crashes in the calling code.\\
To mitigate this risk, the method should catch potential exceptions and provide more informative error messages or handle them gracefully. For example, it could log the exception and return a custom error object instead of returning `null`.\\
\verb|</thinking>|\\
\verb|<reflection>|\\
The method does not handle exceptions properly, which can lead to unexpected behavior or crashes in the calling code. A better approach would be to catch potential exceptions and provide more informative error messages or handle them gracefully.\\
\verb|</reflection>|\\
\verb|<output>|
YES
\verb|</output>|
    \end{minipage} 
    \\
    \multicolumn{1}{c}{\cellcolor{cyan!10} \textbf{\small MSIVD (CodeLlama-13B)}} \\
    \begin{minipage}{\textwidth}
    \vspace{.2em}
    Yes.
\verb|</s>|\\
The vulnerability is type CWE-119, complexity Medium, with the following description: \verb|'''|The vrrp\_print\_json function in vrrp.c in keepalived before 2.1.3 allows remote attackers to cause a denial of service (NULL pointer dereference and daemon crash) or obtain sensitive information from process memory via a crafted JSON file.\verb|'''|
\verb|</s>|\\
    \vspace{.2em}
    \end{minipage} 
    \\
    \hline
    \end{tabular}
    \captionof{table}{Comparison of reasoning output of $\name$ (1.5B), Qwen2.5-Coder-1.5B-Instruct, and MSIVD (CodeLlama-13B) for the vulnerable Java function above.}
    \label{tab:example_java}
    
\end{figure*}

\begin{figure*}[!t]
    \centering
    \includegraphics[width=0.75\linewidth]{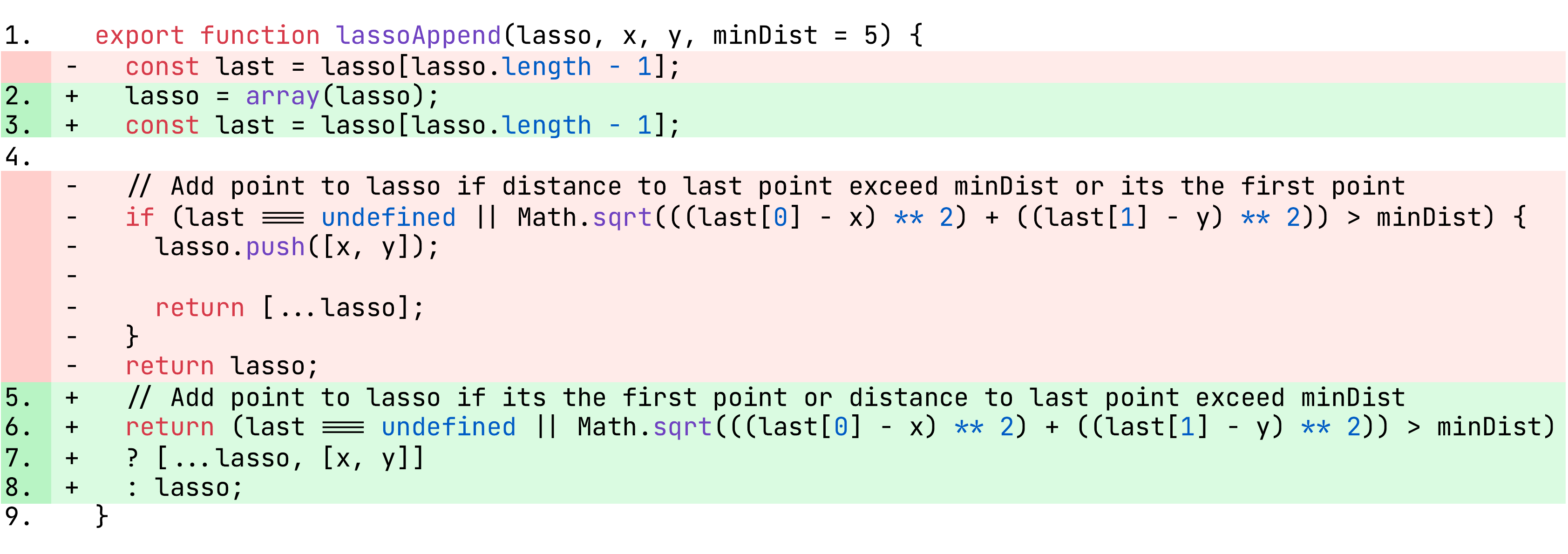}
    \caption{Vulnerable JavaScript function associated with CVE-2023-26487 and CWE-79 (Improper Neutralization of Input During Web Page Generation (Cross-site Scripting)).}
    \label{fig:example_javascript}
    
    \vspace{1em}
    
    \small
    \begin{tabular}{p{\textwidth}}
    \hline
    \multicolumn{1}{c}{\cellcolor{blue!10} \textbf{\small R2Vul 1.5B (ours)}} \\
    \begin{minipage}{\textwidth}
    \vspace{.2em}

\verb|<thinking>|\\
1. **Specific Code Constructs**: The vulnerability primarily stems from the comparison `Math.sqrt(((last[0] - x) ** 2) + ((last[1] - y) ** 2)) $>$ minDist`. This line calculates the Euclidean distance between the current point `(x, y)` and the last point in the `lasso` array. If the calculated distance is greater than `minDist`, the new point `[x, y]` is appended to the `lasso` array.\\

2. **Mechanism of the Vulnerability**: The issue arises because the code does not check if the `lasso` array has more than two elements before attempting to access the second-to-last element (`lasso[lasso.length - 2]`). If the `lasso` array contains only one or zero elements, accessing `lasso[lasso.length - 2]` will result in an out-of-bounds error. This can lead to runtime errors such as `TypeError` or `IndexError`.\\

3. **Potential Impact**: Exploiting this vulnerability could cause the application to crash unexpectedly due to unhandled exceptions. In some cases, depending on how the code handles exceptions, it might also allow attackers to execute arbitrary code or gain unauthorized access. However, the primary impact is likely denial of service (DoS), where the application becomes unavailable until restarted.\\

4. **Contextual Relevance**: This vulnerability aligns with CWE-125, which describes "Out-of-Bounds Read". It occurs when a program reads data past the end of the intended buffer, leading to undefined behavior. In this case, the program attempts to read the second-to-last element of the `lasso` array without verifying that the array contains at least two elements. The CVE-2021-4264 specifically mentions this issue in the context of the Lasso library, indicating that versions prior to 1.3.2 were affected by this exact problem. The vulnerability was addressed in later versions by ensuring proper bounds checking before accessing array elements.\\
\verb|</thinking>|\\
\verb|<output>|
YES
\verb|</output>|

    \end{minipage} 
    \\
    \multicolumn{1}{c}{\cellcolor{cyan!10} \textbf{\small Qwen2.5-Coder-1.5B-Instruct (CoT+Reflection)}} \\
    \begin{minipage}{\textwidth}
    \vspace{.2em}

\verb|<Thinking>|\\
The function `lassoAppend` checks if the new point `(x, y)` should be added to the `lasso` array based on the distance from the last point in the array. However, there is no explicit check to ensure that the `lasso` array is not empty before accessing the last element. This could lead to an error if the `lasso` array is empty when trying to access `lasso[lasso.length - 1]`.

If `lasso` is empty, accessing `lasso[lasso.length - 1]` would throw an `IndexError`. To prevent this, we can add a check at the beginning of the function to ensure that `lasso` is not empty before attempting to access its last element.

Here's the modified function with the added check:
```javascript
\texttt{...}
```

With this modification, the function now includes a check to ensure that `lasso` is not empty before attempting to access its last element. This prevents the `IndexError` and ensures that the function behaves correctly even when the `lasso` array is empty.

\verb|<Reflection>|\\
The original function does not have a check to ensure that `lasso` is not empty before accessing its last element. This could lead to an error if the `lasso` array is empty when trying to access `lasso[lasso.length - 1]`. By adding a check at the beginning of the function, we prevent this potential issue and ensure that the function behaves correctly even when the `lasso` array is empty.

\verb|<output>|
YES
\verb|</output>|

    \end{minipage} 
    \\
    \multicolumn{1}{c}{\cellcolor{cyan!10} \textbf{\small MSIVD (CodeLlama-13B)}} \\
    \begin{minipage}{\textwidth}
    \vspace{.2em}
Yes.
\verb|</s>|\\
The vulnerability is type CWE-119, complexity Medium, with the following description: '''Buffer overflow in the GetVIFFImage function in coders/viff.c in ImageMagick before 6.9.4-5 allows remote attackers to cause a denial of service (application crash) via a crafted file.'''
\verb|</s>|
    \vspace{.2em}
    \end{minipage} 
    \\
    \hline
    \end{tabular}
    \captionof{table}{Comparison of reasoning output of $\name$ (1.5B), Qwen2.5-Coder-1.5B-Instruct, and MSIVD (CodeLlama-13B) for the vulnerable JavaScript function above.}
    \label{tab:example_javascript}
    
\end{figure*}

\begin{figure*}[!t]
    \centering
    \includegraphics[width=0.8\linewidth]{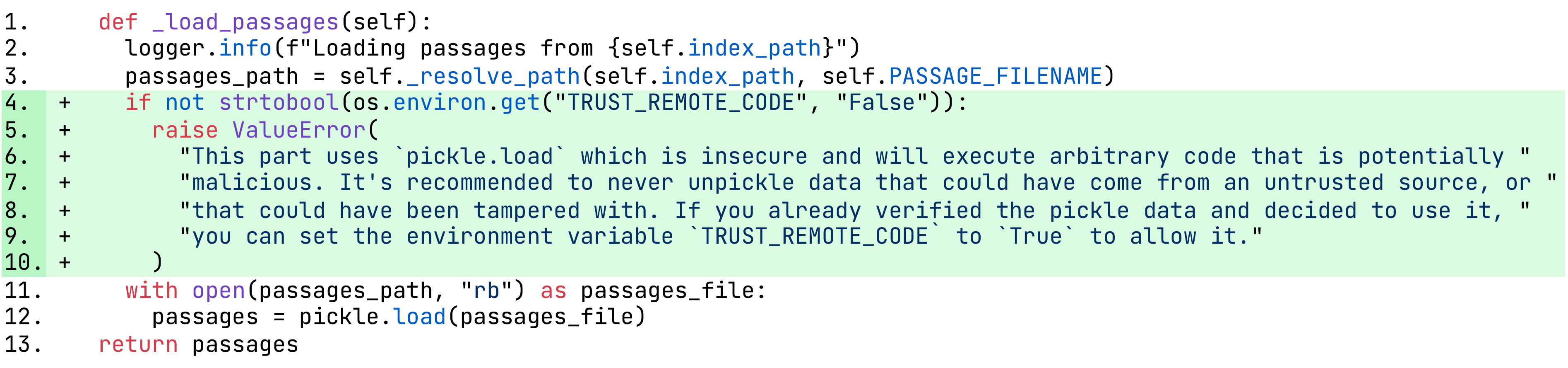}
    \caption{Vulnerable Python function associated with CVE-2023-6730 and CWE-502 (Deserialization of Untrusted Data).}
    \label{fig:example_python}
    
    \vspace{1em}
    
    \small
    \begin{tabular}{p{\textwidth}}
    \hline
    \multicolumn{1}{c}{\cellcolor{blue!10} \textbf{\small R2Vul 1.5B (ours)}} \\
    \begin{minipage}{\textwidth}
    \vspace{.2em}

\verb|<thinking>|\\
1. **Specific Code Constructs**: The specific code construct contributing to the vulnerability is the call to `pickle.load(passages\_file)` within the `\_load\_passages` function. This line deserializes data from a file into Python objects without any form of validation or sanitization.\\

2. **Mechanism of the Vulnerability**: The mechanism of the vulnerability lies in the use of Python's `pickle` module for deserialization. The `pickle.load()` function can execute arbitrary code during the deserialization process if the serialized data includes malicious payloads. In the context of CVE-2023-48795, this means that an attacker could craft a `.pkl` file containing malicious code that, when loaded by the `pickle.load()` function, would execute arbitrary commands on the system. This is possible because the `pickle` module does not perform any checks on the content of the file before executing it.\\

3. **Potential Impact**: Exploiting this vulnerability could allow an attacker to execute arbitrary code with the privileges of the process running the `\_load\_passages` function. This could lead to unauthorized access, data theft, system compromise, or even complete control over the affected system. Given that the CVE description mentions that the attack vector involves uploading a `.pkl` file, the immediate risk is to upload a malicious file that upon loading causes harm.\\

4. **Contextual Relevance**: The vulnerability is directly related to CWE-502, which describes Deserialization of Untrusted Data. This CWE highlights the risks associated with deserializing data from untrusted sources without proper validation, which can lead to remote code execution. Additionally, CWE-138, Improper Neutralization of Special Elements used in an OS Command ('OS Command Injection'), although not explicitly mentioned in the CVE description, shares the theme of improper handling of external inputs leading to command injection attacks. However, in this case, the primary concern is the deserialization aspect rather than command injection. The CVE-2023-48795 specifically addresses the issue in the context of the Hugging Face Transformers library, where the `pickle` module was used unsafely for model evaluation, aligning with the identified vulnerability in the provided code snippet.\\
\verb|</thinking>|\\
\verb|<output>|
YES
\verb|</output>|

    \end{minipage} 
    \\
    \multicolumn{1}{c}{\cellcolor{cyan!10} \textbf{\small Qwen2.5-Coder-1.5B-Instruct (CoT+Reflection)}} \\
    \begin{minipage}{\textwidth}
    \vspace{.2em}

\verb|<Thinking>|\\
The given function `\_load\_passages` appears to be designed to load passages from a file using Python's `pickle` module. However, there are several potential vulnerabilities that could be present in this function:

1. **File Path Injection**: The function uses the `self.index\_path` variable to construct the path to the passage file. If `self.index\_path` is controlled by an attacker, an attacker could manipulate it to point to a malicious location, potentially leading to arbitrary code execution or other security issues.

2. **Untrusted Input**: The function does not validate or sanitize the input parameters before using them in the file path construction. This could lead to directory traversal attacks if the `self.index\_path` is not properly validated.

3. **Resource Leaks**: The function opens the file in binary mode (`"rb"`), which can consume significant resources if the file is large. Additionally, the file is loaded into memory using `pickle.load`, which can also consume significant memory if the file is large.

4. **Error Handling**: While the function includes basic logging, it does not handle exceptions that might occur during file operations or unpickling. This could lead to silent failures or unexpected behavior.

5. **Security Considerations**: There are no additional security considerations mentioned in the function, such as encryption or access controls.

6. **Code Complexity**: The function is relatively simple and straightforward, but it still contains some potential vulnerabilities.

\verb|<Reflection>|
The function does contain several potential vulnerabilities. Specifically, the file path is constructed using `self.index\_path`, which could be manipulated by an attacker to point to a malicious location. Additionally, the function does not validate or sanitize the input parameters before using them in the file path construction, which could lead to directory traversal attacks. Furthermore, the function does not handle exceptions that might occur during file operations or unpickling, which could lead to silent failures or unexpected behavior.

\verb|<output>|
YES
\verb|</output>|

    \end{minipage} 
    \\
    \multicolumn{1}{c}{\cellcolor{cyan!10} \textbf{\small MSIVD (CodeLlama-13B)}} \\
    \begin{minipage}{\textwidth}
    \vspace{.2em}
Yes.
\verb|</s>|\\
The vulnerability is type CWE-125, complexity Medium, with the following description: '''The Juniper parser in tcpdump before 4.9.2 has a buffer over-read in print-juniper.c allows remote attackers to cause a denial of service (application crash) or obtain sensitive information from process memory via a crafted packet.'''
\verb|</s>|
    \vspace{.2em}
    \end{minipage} 
    \\
    \hline
    \end{tabular}
    \captionof{table}{Comparison of reasoning output of $\name$ (1.5B), Qwen2.5-Coder-1.5B-Instruct, and MSIVD (CodeLlama-13B) for the vulnerable Python function above.}
    \label{tab:example_python}
    
\end{figure*}

We illustrate examples of reasoning output on all five languages in Figures~\ref{fig:example_c_sharp}--\ref{fig:example_c} and Tables~\ref{tab:example_c_sharp}--\ref{tab:example_c} for three models: $\name$ 1.5B, Qwen2.5-Coder-1.5B-Instruct (Qwen2.5-1.5B) with CoT and a reflection step, and MSIVD using CodeLlama-13B.
Across all examples, we observe that MSIVD's outputs are erroneous. 
Therefore, we focus the following comparison on $\name$ and Qwen2.5-1.5B.

\subsubsection{C\# Example (Table~\ref{tab:example_c_sharp}).} 
$\name$ demonstrates security awareness by identifying the problematic use of \texttt{Uri.EscapeUriString()} in URL construction and recognizing this as a potential injection vector.
In contrast, Qwen2.5-1.5B fails to identify any vulnerability, dismissing the security concern due to the presence of URI escaping without recognizing that \texttt{Uri.EscapeUriString()} is insufficient for preventing URL-based attacks. 
$\name$'s recognition of URL construction as an attack surface makes it substantially more accurate than Qwen2.5-1.5B's complete miss of the security issue.

\subsubsection{Java Example (Table~\ref{tab:example_java}).}

$\name$ demonstrates strong security awareness by correctly identifying the path traversal vulnerability, explaining how unsanitized user input in the \texttt{name} parameter can be exploited through directory traversal sequences to create directories in unauthorized locations, and properly mapping it to CWE-22. 
In contrast, Qwen2.5-1.5B entirely misses the security vulnerability, focusing on benign code quality issues like exception handling and null returns, which have no security implications. 

\subsubsection{JavaScript Example (Table~\ref{tab:example_javascript}).}

Both reasonings fundamentally misunderstand the vulnerability. 
$\name$ incorrectly identifies an out-of-bounds read issue with \texttt{lasso[lasso.length - 2]} (which doesn't exist in the code), while Qwen2.5-1.5B misidentifies a potential empty array access issue with \texttt{lasso[lasso.length - 1]}. 
Neither recognizes the actual type confusion vulnerability where an attacker can substitute the lasso array with a malicious object containing a weaponized push method, enabling XSS attacks. 
The patch's addition of \texttt{array(lasso)} validates the input type to prevent this injection vector, which both reasonings miss.

\subsubsection{Python Example (Table~\ref{tab:example_python}).}

$\name$ demonstrates precise security knowledge by correctly identifying the pickle deserialization vulnerability as the primary security concern, accurately explaining how \texttt{pickle.load()} can execute arbitrary code embedded in malicious \texttt{.pkl} files, and properly mapping them to CWE-502. 
In contrast, Qwen2.5-1.5B, despite concluding that a vulnerability exists, misidentifies the threat vector by focusing on file path manipulation and resource consumption issues while completely overlooking the fundamental insecurity of unpickling untrusted data.

\begin{figure*}[!t]
    \centering
    \includegraphics[width=0.65\linewidth]{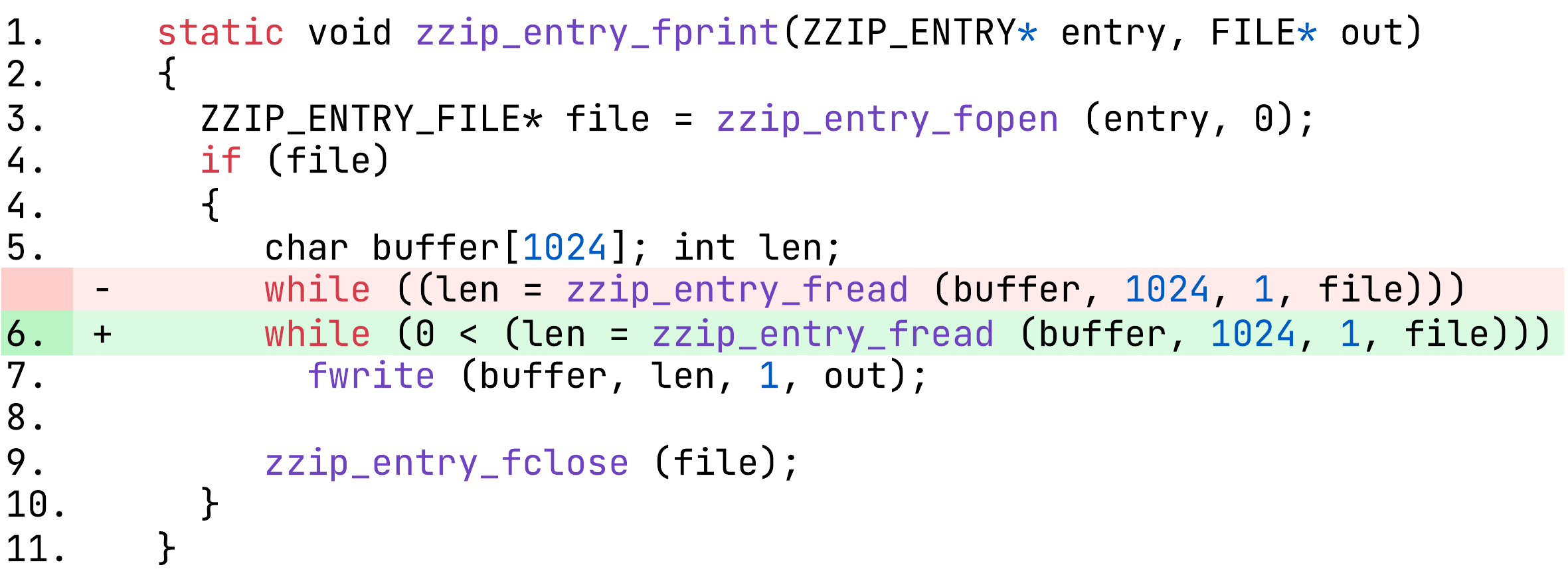}
    \caption{Vulnerable Python function associated with CVE-2020-18442 and CWE-835 (Loop with Unreachable Exit Condition (Infinite Loop)).}
    \label{fig:example_c}
    
    \vspace{1em}
    
    \small
    \begin{tabular}{p{\textwidth}}
    \hline
    \multicolumn{1}{c}{\cellcolor{blue!10} \textbf{\small R2Vul 1.5B (ours)}} \\
    \begin{minipage}{\textwidth}
    \vspace{.2em}
\verb|<thinking>|\\
1. **Specific Code Constructs**: The vulnerability arises from the `fwrite` call within the loop. Specifically, the line `fwrite (buffer, len, 1, out);` is problematic because it writes data without checking if there is enough space in the output buffer (`out`) to accommodate the data being written.\\

2. **Mechanism of the Vulnerability**: The issue stems from the fact that `fwrite` does not perform bounds checking on the destination buffer size. If `len` exceeds the current size of the output buffer pointed to by `out`, `fwrite` will attempt to write beyond the allocated memory, leading to undefined behavior. This can result in a heap-based buffer overflow, where an attacker can overwrite adjacent memory, potentially altering program execution flow or corrupting data structures.\\

3. **Potential Impact**: Exploiting this vulnerability could allow an attacker to execute arbitrary code, cause a denial of service by crashing the application, or leak sensitive information. By manipulating the contents of the input data or the state of the application, an attacker might be able to bypass security checks or gain unauthorized access.\\

4. **Contextual Relevance**: This vulnerability aligns with CWE-787, which describes Out-of-bounds Write. It specifically matches CVE-2019-16515, indicating that the flaw occurs in the `zzip\_entry\_fprint()` function due to improper handling of the `fwrite` operation. The CVE description mentions that the vulnerability allows attackers to execute arbitrary code, crash the application, or leak sensitive information, which directly corresponds to the potential impacts discussed above.\\
\verb|</thinking>|\\
\verb|<output>|
YES
\verb|</output>|

    \end{minipage} 
    \\
    \multicolumn{1}{c}{\cellcolor{cyan!10} \textbf{\small Qwen2.5-Coder-1.5B-Instruct (CoT+Reflection)}} \\
    \begin{minipage}{\textwidth}
    \vspace{.2em}
\verb|<Thinking>|\\
The given function `zzip\_entry\_fprint` appears to be designed to print the contents of a ZIP entry to a file. However, there are a few potential vulnerabilities that could be exploited:

1. **Buffer Overflow**: The function reads up to 1024 bytes from the ZIP entry using `zzip\_entry\_fread`. If the ZIP entry contains more data than this buffer can hold, it may lead to a buffer overflow. This could potentially allow an attacker to overwrite memory locations beyond the intended buffer size, leading to arbitrary code execution.

2. **File Handling Issues**: The function does not handle file opening and closing properly. If the file cannot be opened, the function will continue to attempt to read from it, which could lead to undefined behavior or crashes.

3. **Resource Management**: There is no explicit resource management in place for the ZIP entry file. If the function is called multiple times without closing the file after each call, it could lead to resource leaks.

4. **Error Handling**: The function does not provide any error handling for file operations. If the file cannot be opened or read, the function will continue to execute, which could lead to unexpected behavior or crashes.

5. **Security Considerations**: The function does not perform any security checks on the input parameters, such as ensuring that the ZIP entry pointer is valid or that the file pointer is not null.\\

\verb|<Reflection>|
The function does contain some potential vulnerabilities related to buffer overflows and improper file handling. However, without additional context about the specific implementation and usage of the `zzip\_entry\_fprint` function, it is difficult to determine the exact impact of these vulnerabilities.

\verb|<output>|
YES
\verb|</output>|

    \end{minipage} 
    \\
    \multicolumn{1}{c}{\cellcolor{cyan!10} \textbf{\small MSIVD (CodeLlama-13B)}} \\
    \begin{minipage}{\textwidth}
    \vspace{.2em}
Yes.
\verb|</s>|\\
The vulnerability is type CWE-190, complexity Medium, with the following description: '''Integer overflow in the zzip\_entry\_fread function in zzip.c in libzzip before 0.10 allows remote attackers to cause a denial of service (application crash) via a crafted ZIP archive.'''
\verb|</s>|
    \vspace{.2em}
    \end{minipage} 
    \\
    \hline
    \end{tabular}
    \captionof{table}{Comparison of reasoning output of $\name$ (1.5B), Qwen2.5-Coder-1.5B-Instruct, and MSIVD (CodeLlama-13B) for the vulnerable C function above.}
    \label{tab:example_c}
    
\end{figure*}

\subsubsection{C Example (Table~\ref{tab:example_c}).}
Both reasonings miss the actual infinite loop vulnerability identified in CVE-2020-18442, though $\name$ is marginally closer by at least focusing on a specific line of code that's part of the vulnerable construct.
$\name$ incorrectly focuses on buffer overflow risks in \texttt{fwrite}, erroneously claiming it lacks bounds checking on the output buffer and could lead to arbitrary code execution. Qwen2.5-1.5B provides a generic list of potential issues, including buffer overflows and file handling problems, without identifying any specific vulnerability mechanism. 
Neither reasoning recognizes that the actual issue is the loop condition failing to properly handle error returns from \texttt{zzip\_entry\_fread}, where negative return values would satisfy the original while condition, causing an infinite loop. While both are incorrect, $\name$'s attempt to analyze the specific \texttt{fwrite} operation within the vulnerable loop shows slightly better focus than Qwen2.5-1.5B's scattered approach.

These qualitative examples reveal that our preference tuning approach improves vulnerability detection capabilities and reasoning, with $\name$ demonstrating more precise identification of security-critical code patterns, though the persistent failure cases across both models underscore the complexity of comprehensive vulnerability reasoning.

\subsection{Prompt Templates}
\label{sec_ap:prompts}

\begin{itemize}
    \item Prompt used for re-labelling our raw dataset: Table~\ref{tab:apx_prompt_labelling}.

    \item Prompt for generating structured reasoning: Table~\ref{tab:apx_prompt_generation}.

    \item Prompt for CoT, SFT, and $\name$ inference: Table~\ref{tab:apx_prompt_cot}.

    \item Prompt for qualitative reasoning assessment using LLM-as-a-judge: Table~\ref{tab:apx_prompt_qualitative}.
\end{itemize}

\begin{table*}[!t]
\centering
\small

\begin{tabular}{p{15cm}}
    \toprule
Please evaluate the code changes on a scale from 0 to 4 based on the following criteria:\\
- Score 0: The changes are unrelated to fixing security vulnerabilities.\\
- Score 1: The changes may involve minor refactoring, but do not focus on fixing security vulnerabilities.\\
- Score 2: The changes suggest some improvement in code quality, with potential indirect effects on security, but not directly targeting vulnerabilities.\\
- Score 3: The changes have a moderate focus on addressing security vulnerabilities.\\
- Score 4: The changes clearly and directly target security vulnerabilities, providing a strong fix.\\
\\
The length of the code should not influence your evaluation. Focus on the logic of the changes, line by line, and their relevance to vulnerability fixing.
Consider the context provided by other functions in the same commit and the information about the CVE linked to this commit. Think step by step.\\
\\
\#\#\# CVE ID:\\
\verb|{cve_id}| \\
\\
\#\#\# CVE Description:\\
\verb|{cve_description}| \\
\\
\#\#\# Commit Message:\\
\verb|{commit_message}| \\
\\
\#\#\# Original code snippet (code before changes):\\
\verb|{original_code}| \\
\\
\#\#\# Revised code snippet (code after changes):\\
\verb|{revised_code}| \\
    \bottomrule
\end{tabular}
\caption{Prompt used to re-label our dataset using GPT-4o (adapted from~\cite{li2024cleanvul}).}
\label{tab:apx_prompt_labelling}
\end{table*}

\begin{table*}[!t]
\centering
\small
\begin{tabular}{p{15cm}}
    \toprule
    \multicolumn{1}{c}{Prompt for vulnerable functions} \\
    \midrule
    The following function has been flagged as vulnerable. 

Input function: \\
\verb|```{lang}| \\
\verb|{function}| \\
\verb|```|

This function contains a vulnerability associated with the following CWE(s): \verb|{cwe_list}|. 
Specifically, it is linked to \verb|{cve_id}|, which is described as follows: 
\verb|{cve_desc}|.

Given this information, generate a detailed and coherent thought process within the \verb|<thinking>| tags. Your reasoning should focus on the following elements: \\
1. \textbf{Specific Code Constructs}: Identify the parts of the code that directly contribute to the vulnerability. \\
2. \textbf{Mechanism of the Vulnerability}: Explain how the identified code leads to the vulnerability (e.g., unsafe function calls, lack of input validation). \\
3. \textbf{Potential Impact}: Describe the consequences of exploiting this vulnerability. \\
4. \textbf{Contextual Relevance}: Relate your explanation to the provided CWE(s) and CVE description.
Strictly follow these steps in your reasoning. Do not include more steps in your reasoning. \\
    \midrule
    \multicolumn{1}{c}{Prompt for non-vulnerable functions} \\
    \midrule
    The following function has been flagged as non-vulnerable. 

Input function: \\
\verb|```{lang}| \\
\verb|{function}| \\
\verb|```|

This function has been reviewed and determined to not contain any known vulnerabilities. 

Given this information, generate a detailed and coherent thought process within the \verb|<thinking>| tags. Your reasoning should focus on the following elements: \\
1. \textbf{Analysis of Code Safety}: Identify specific aspects of the code that contribute to its security, such as proper use of safe coding practices or robust validation mechanisms. \\
2. \textbf{Absence of Common Vulnerabilities}: Discuss potential vulnerabilities that could arise in similar functions and explain why they are not applicable here. \\
3. \textbf{Validation of the Non-Vulnerable Label}: Provide evidence-based reasoning to justify why the function is secure and free of exploitable flaws.
Strictly follow these steps in your reasoning. Do not include more steps in your reasoning. \\
    \bottomrule
\end{tabular}
\caption{Prompt templates used to generate structured reasoning for vulnerable and non-vulnerable functions.}
\label{tab:apx_prompt_generation}
\end{table*}

\begin{table*}[!t]
\centering
\small
\begin{tabular}{p{15cm}}
    \toprule
    \multicolumn{1}{c}{System prompt} \\
    \midrule
    You are an AI assistant expert in software vulnerabilities. You use a chain-of-thought approach to answer queries.Follow these steps:\\
    1. Think through the problem step by step within the \verb|<thinking>| tags.\\
    2. Provide your final, concise answer within the \verb|<output>| tags.\\
\\
    Important: The \verb|<thinking>| section is for your internal reasoning process only.\\
    Do not include any part of the final answer in this section.\\
    The actual response for the query must be entirely contained within the \verb|<output>| tags.\\
\\
Strictly use the following format for your response:\\
\verb|<thinking>| \\
\verb|[|Your step-by-step reasoning goes here. This is your internal thought process, not the final answer.\verb|]|\\
\verb|</thinking>| \\
\verb|<output>| \\
\verb|[|Your final, concise answer to the query. This is the only part that will be shown to the user.\verb|]|
\verb|</output>| \\
    \midrule
    \multicolumn{1}{c}{User prompt} \\
    \midrule
    Analyze the following function and determine whether it contains any vulnerabilities. Indicate your final decision with:\\
    - YES: the function is vulnerable\\
    - NO: the function is not vulnerable\\
Output your final decision within the \verb|<output>| and \verb|</output>| tags. Strictly follow the output format.\\
\\
Input function: \\
\verb|```{lang}| \\
\verb|{function}| \\
\verb|```| \\
    \bottomrule
\end{tabular}
\caption{System and user prompts used for CoT, SFT, and R2Vul.}
\label{tab:apx_prompt_cot}
\end{table*}

\begin{table*}[!t]
\centering
\small
\begin{tabular}{p{15cm}}
    \toprule
    \multicolumn{1}{c}{System prompt} \\
    \midrule
    You are an expert in software security and vulnerability detection. Your task is to evaluate the reasoning outputs from three language models for a given code function. \\
    \midrule
    \multicolumn{1}{c}{User prompt} \\
    \midrule
    **Code Function:**\\
\verb|{function}| \\
\\
**Ground-Truth Label:** \\
\verb|{label}| \\
\\
**Reasoning Outputs:** \\
1. Reasoning A: \\
\verb|{reasoning_a}| \\
\\
2. Reasoning B: \\
\verb|{reasoning_b}| \\
\\
3. Reasoning C: \\
\verb|{reasoning_c}| \\
\\
**Evaluation Instructions:**\\
1. **Comparative Analysis:** Compare the three reasoning outputs in terms of correctness, completeness, clarity, and actionability. Highlight strengths and weaknesses of each reasoning output relative to one another. \\
2. **Score Assignment:** After comparing, score each reasoning output on a scale of 1 to 5 for the following criteria: \\
   - Completeness: Does the reasoning cover the vulnerability mechanism or justify why the code is safe, considering edge cases and attack vectors? \\
   - Clarity: Is the reasoning logically structured, free of ambiguities, and using precise technical terms? \\
   - Actionability: Does the reasoning provide actionable insights like highlighting vulnerable lines, suggesting patches, or detailing risks? \\
3. **Final Selection:** Conclude by selecting the best reasoning output overall, justifying why it stands out based on the comparison and scores. \\
    \bottomrule
\end{tabular}
\caption{System and user prompts used in the qualitative assessment of reasoning experiments.}
\label{tab:apx_prompt_qualitative}
\end{table*}

\end{document}